\documentclass[11pt,a4paper]{article}
\usepackage{jcappub}
\usepackage{bm}
\usepackage{amsmath}
\usepackage{latexsym,amsmath,amssymb,amsfonts,mathrsfs}
\usepackage{appendix}
\usepackage[usenames,dvipsnames]{xcolor}
\usepackage{verbatim}
\usepackage{ulem}

\def\dd{\mathrm{d}}
\def\mcA{\mathcal{A}}
\def\mcP{\mathcal{P}}
\def\em{{\rm em}}
\def\inf{{\rm inf}}
\def\tot{{\rm tot}}
\def\obs{{\rm obs}}

\def\Mpl{M_{\rm Pl}}
\def\GeV{{\rm GeV}}
\def\Mpc{{\rm Mpc}}

\def\G{{\rm G}}

\def\osc{{\rm osc}}
\def\BD{{\rm BD}}

\title{
Pre-reheating Magnetogenesis 
in the Kinetic Coupling Model 
}
\author[a]{Tomohiro Fujita}
\author[b]{, Ryo Namba}

\affiliation[a]{Stanford Institute for Theoretical Physics and Department of Physics,\\ 
Stanford University, Stanford, CA 94306, USA}
\affiliation[b]{Kavli Institute for the Physics and Mathematics of the
Universe (WPI), TODIAS,\\  
the University of Tokyo, 5-1-5
Kashiwanoha, Kashiwa, 277-8583, Japan}

\emailAdd{tomofuji@stanford.edu}
\emailAdd{ryo.namba@ipmu.jp}

\abstract{
Recent blazar observations provide growing evidence for the presence of magnetic fields in the extragalactic regions.
While a natural speculation is to associate the production to inflationary physics, it has been known that magnetogenesis solely from inflation is quite challenging. 
We therefore study a model in which a non-inflaton field $\chi$ coupled to the electromagnetic field through its kinetic term, $-I^2(\chi) F^2 /4$, continues to move after inflation until the completion of reheating. This leads to a post-inflationary amplification of the electromagnetic field.
We compute all the relevant contributions to the curvature perturbation, including gravitational interactions, and impose the constraints from the CMB scalar fluctuations on the strength of magnetic fields. 
We, for the first time, explicitly verify both the backreaction and CMB constraints in a simple yet successful magnetogenesis scenario without invoking a dedicated low-scale inflationary model in the weak-coupling regime of the kinetic coupling model.
}

\arxivnumber{1602.05673}

\usepackage{color}

\begin{document}

\begin{flushright}
IPMU16-0017
\end{flushright}

\maketitle

%
%
%
\section{Introduction}

Observations have revealed that our universe is magnetized on various different scales. One of the most intriguing scales is the largest one. It is known that galaxies and their clusters have their own magnetic fields with     the typical strength $\mathcal{O}(10^{-6}) \, \G$. However, their origin is a long-term open question. Furthermore, the multi-frequency blazar observation implies that the magnetic fields which are not associated with galaxies or clusters do exist~\cite{Neronov:1900zz, Tavecchio:2010mk, Dolag:2010ni, Essey:2010nd, Taylor:2011bn, Takahashi:2013uoa, Chen:2014rsa}.
They are called extragalactic magnetic fields (EGMF) or void magnetic fields, and their strength are estimated to be no less than $\mathcal{O}(10^{-15}) \, \G$~\cite{Taylor:2011bn}.%
\footnote{
The value of the lower bound on the EGMF strength varies by one or two orders of magnitude (roughly ${\cal O}(10^{-17}) - {\cal O}(10^{-15}$) depending on assumptions in the analysis, such as the energy spectrum and time variation of the source. In this paper, we aim at generation of more sufficient, i.e. larger, magnetic fields and consider the upper value, ${\cal O}(10^{-15}) \, \G$.
}
EGMF may also indicate that the galaxy and cluster magnetic fields have a cosmological origin. Nevertheless,
it is difficult even to find a hypothetical scenario which explains the origin of EGMF in a consistent and quantitative way without fine tuning, and thus EGMF has attracted attention as a unique arena of theoretical cosmology and astrophysics (for recent review see \cite{Kandus:2010nw, Durrer:2013pga, Subramanian:2015lua}). The blazar observations put the lower bound not on the strength of EGMF itself but on the following {\it effective} strength of EGMF~\cite{Fujita:2012rb,Caprini:2015gga}, i.e.,
\begin{align}
B_{\rm eff} \gtrsim 10^{-15}\G
\label{Target}
\end{align}
with
\begin{align}
 B_{\rm eff}^2(\eta_{\rm now})
  \equiv \int^{k_{\rm diff}}_0 \frac{{\rm d} k}{k} 
  F(kL)\, \mathcal{P}_{B}(\eta_{\rm now},k),
 \label{Beff}
  \\
 F(z) \equiv \frac{3}{2}z^{-2}
  \left[ \cos (z) - \frac{\sin (z)}{z} + z {\rm Si}(z) \right]
 \label{Fz-def}
  .
\end{align}
Here $\mathcal{P}_{B}(\eta_{\rm now},k)$ is the power spectrum of EGMF at present, Si$(z)$ denotes the sine integral function, $k_{\rm diff}^{-1}\sim 100$AU is the wave number corresponding to the present cosmic diffusion length, and $L \simeq$ 1Mpc stands for the characteristic length scale for energy
losses of charged particles due to inverse Compton scattering.
Since $F(kL) \propto k^{-1}$ for $k\gtrsim L^{-1}$ and it suppresses the contribution from smaller scales than $L$, it is favorable to produce large-scale magnetic fields to explain the blazar observation.

The magnetic fields present in the line of sight of the blazar photons are in the extragalactic regions, and hence astrophysical processes are hardly responsible for their generation. 
A compelling possibility is to attribute them to a cosmological origin. There have been dedicated studies on several different mechanisms to produce large-scale magnetic fields in those regions. As a small subset of examples, collision of bubbles created at a phase transition in the early universe, such as QCD and electroweak, can produce magnetic fields~\cite{Vachaspati:1991nm,Sigl:1996dm}. A concrete model that quantitatively account for the blazar observations is, however, not yet well established. Also by the second-order perturbation theory in the plasma, the effective difference in the motion of charged particles can induce magnetic fields in cosmological scales~\cite{Matarrese:2004kq,Saga:2015bna}. Quantitative studies have shown that the effective magnetic strength from the second-order effects does not reach the observed value.

Inflationary magnetogenesis, i.e.~the generation of magnetic fields during inflation, has been intensively investigated~\cite{Turner:1987bw,Ratra:1991bn,Garretson:1992vt,
Finelli:2000sh,Davis:2000zp,Bamba:2003av,Anber:2006xt,Martin:2007ue,Durrer:2010mq,Ferreira:2013sqa,
Caprini:2014mja,Kobayashi:2014sga,Tasinato:2014fia,Fujita:2015iga,Domenech:2015zzi,Campanelli:2015jfa}.%
\footnote{Magnetogenesis in bouncing universe scenarios is considered in \cite{Salim:2006nw,Membiela:2013cea,Sriramkumar:2015yza}.}
This is because large-scale structures are believed to be seeded in the inflationary era and the idea is naturally extended to explore the similar possibility for magnetic fields on large scales. The most well-studied model of inflationary magnetogenesis is the kinetic coupling model (a.k.a. $I^2FF$ model) proposed by Ratra~\cite{Ratra:1991bn}. In this model, a rolling scalar field is coupled with the kinetic term of the gauge field, and the energy density of the scalar field is transferred to the electromagnetic sector.
Another model in which a rolling pseudo-scalar field drives magnetogenesis is also well studied~\cite{Garretson:1992vt,Anber:2006xt,Durrer:2010mq,Caprini:2014mja, Fujita:2015iga}.
Although quite a few models have been proposed and explored so far, each of them has to face all of the following problems~\cite{Fujita:2012rb, Demozzi:2009fu, Barnaby:2012xt, Fujita:2013qxa, Fujita:2014sna, Ferreira:2014hma,Ferreira:2014zia,Ferreira:2014mwa}: {\it (i) The backreaction problem}: the energy density of the generated electromagnetic fields must not exceed the inflaton energy density during inflation. {\it (ii) The strong coupling problem}: the effective coupling constant between the gauge field and charged fermions should be small to verify the perturbative calculation. {\it (iii) The curvature perturbation problem}: the curvature perturbation induced by the electromagnetic fields must be consistent with CMB observations.
It has been pointed out that imposing the conditions to resolve these issues (i)-(iii) is quite challenging and that primordial magnetogenesis solely from inflation at least requires a dedicated low-energy inflationary model, whose energy scale is close to the threshold of Big Bang nucleosynthesis (BBN)~ \cite{Fujita:2014sna,Ferreira:2014hma}.

In this paper, we consider a magnetogenesis scenario in the framework of the kinetic coupling model. To overcome the above three problems, we extend the original model in the following way. In the original paper, the scalar field coupled with the electromagnetism is the inflation. Furthermore, it has been often assumed that the kinetic function $I$ which is multiplied by the kinetic term of the gauge field $F_{\mu\nu} F^{\mu\nu}$ is just a simple power-law function of the scale factor, namely $I(t) \propto a^{-n}$, and it varies only during inflation.%
\footnote{The reader should be referred to some important exceptional works~\cite{Bamba:2003av,Ferreira:2013sqa,Kobayashi:2014sga}.} 
However, provided that the kinetic function $I$ is driven by a spectator scalar field which is not the inflaton, it is quite natural that $I$ continues to vary after inflation ends. Thus we assume that $I$ varies until reheating is completed.%
\footnote{Clearly, the spectator field responsible for the time dependence of $I$ does not necessarily decay at the same time as the inflaton, which is the dominant energy content at the time and drives reheating. We simply impose their simultaneous decay as an additional assumption, in order to reduce the number of model parameters.}
Moreover, we also consider that $I$ starts varying after perturbations with the  wave numbers corresponding to the CMB scales exit the horizon to optimize the scenario for magnetogenesis.

In our scenario, since the kinetic coupling is always no less than unity, $I\ge1$, we do not have to be concerned with the strong coupling problem.%
\footnote{A recent study \cite{Domenech:2015zzi} considers an opposite regime, i.e.~$I \le 1$ during inflation, while keeping weak couplings to fermions by introducing a coupling function in the fermion sector as well and by explicitly breaking the $U(1)$ gauge invariance.}
Yet, we need to properly analyze the perturbations of the fields to address the other two problems. We obtain the exact solution of the gauge field and rigourously estimate its energy density and the curvature perturbation induced by it. Furthermore, the curvature perturbation is also produced from the scalar field perturbations which are sourced by the electromagnetic fields through both the direct coupling  and the gravitational interactions. We calculate all of the leading-order contributions and find the constraints on the produced magnetic fields. 
Our result shows that the magnetic fields generated in our scenario can be strong enough to explain the observational value.

This paper is organized as follows. In sec.~\ref{Magnetogenesis}, we explain the setup of our scenario, calculate the evolution of the electromagnetic fields, and obtain the magnetic power spectrum at present. 
In sec.~\ref{Constraints}, the constraints from the backreaction and the induced curvature perturbation are derived. We also make a comment on the interaction between the electromagnetic fields and charged particles.
In sec.~\ref{Results}, the results of this paper are shown. 
Section \ref{Summary} is devoted to the conclusion. 
In appendices, the explicit derivation and expressions of the exact solution of the electromagnetic fields are shown, and the calculation of curvature perturbation is described.

\section{Magnetogenesis}
\label{Magnetogenesis}

\subsection{Model setup}

We consider the kinetic coupling model with the following action:
\begin{equation}
S= \int \dd^4 x \sqrt{-g}
\left[
\frac{M_{\rm Pl}^2}{2} R
- \frac{1}{2}(\partial \phi)^2-V(\phi)
-\frac{1}{2}(\partial\chi)^2-U(\chi)
-\frac{1}{4} I^2(\chi) F_{\mu\nu}F^{\mu\nu}
\right] \; ,
\label{Model Action}
\end{equation}
where $\phi$ is the inflaton, $\chi$ is a spectator scalar field which 
drives the kinetic function $I(\chi)$,
$F_{\mu\nu}\equiv \partial_\mu A_\nu - \partial_\nu A_\mu$,
$A_\mu$ is the U(1) gauge field associated to the electromagnetism, namely the photon,
and $R$ and $M_{\rm Pl}$ are the Ricci scalar and the reduced Planck mass, respectively.
The energy of the background $\chi$ field is transferred to the electromagnetic fields through the kinetic coupling, $I^2 FF$, and thus
the electromagnetic fields are generated in this model.
In this paper, we 
assume that the inflaton oscillation after inflation can be well approximated by the one with a quadratic potential, while we let the explicit forms of $V(\phi)$ during inflation and $U(\chi)$ unspecified and assume simple time evolution of the background universe and of $I(\chi)$.

\begin{figure}[tbp]
  \begin{center}
  \includegraphics[width=75mm]{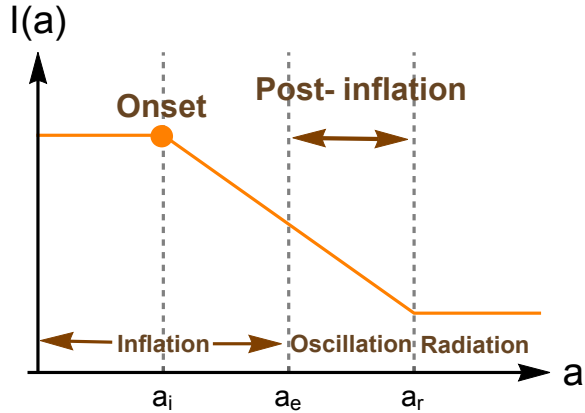}
  \end{center}
  \caption
 {The behavior of $I(a)$ given in eq.~\eqref{simple I}.
}
 \label{simple case}
\end{figure}

We consider the quasi-de Sitter expansion during inflation (i.e.~$H_\inf \approx {\rm const.}$ and $\eta=-1/a H_\inf$ where $a$, $H$ and $\eta$ are the scale factor, Hubble parameter and conformal time, respectively), and the expansion of the matter-dominated universe, $\rho_\phi \propto a^{-3}$ and $\eta=2/aH\propto a^{1/2}$, during the inflaton oscillating phase between the end of inflation and the completion of reheating. 
We  impose that the $\chi$ field is in a perturbative regime, and such time dependence is driven by the homogeneous vacuum expectation value of $\chi$, i.e.~$I(\chi) \approx I(\langle \chi \rangle) = I(a)$.
We assume that $I$ is constant at the beginning, starts varying at a certain time during inflation and ceases to evolve at the completion of reheating. Without loss of generality, we set $I=1$ when it stops.
While it is natural for $\chi$ and thus $I(\chi)$ to evolve after inflation as $\chi$ is not the inflaton but a spectator field, they could have different time dependences during and after inflation and could stop varying at an arbitrary time. We place additional assumptions that $I \propto a^{-n}$ both during and after inflation and $\chi$ decays at the time of reheating, simply to reduce the number of model parameters.
In summary, the behaviors of the background expansion and the kinetic function $I$ are given by
\begin{equation}
\eta = 
\left\{
 \begin{array}{lc}
 -1/a H_\inf &  (a <a_e)\\
 2/aH &  (a_e <a <a_r)
 \end{array} 
 \right.,
 \qquad
I(a) = 
\left\{
 \begin{array}{lc}
 (a_i/a_r)^{-n}\equiv I_i & (a <a_i) \\
 (a/a_r)^{-n} &  (a_i <a <a_r) \\
 1  & \quad (a_r<a) \\
 \end{array} 
 \right. ,
\label{simple I}
\end{equation}
where $a_i, a_e$ and $a_r$ denote the values of scale factor $a$ when $I(a)$ starts varying, inflation ends and reheating completes, respectively.
Fig.~\ref{simple case} illustrates this behavior of $I$.
Since we discuss only the case with $n>5/2$, 
for the reason mentioned in Subsection \ref{subsec:strength-present},
$I(a)$ is always larger than unity and hence our scenario is free from the strong coupling problem, namely the effective electromagnetic coupling strength $e / I < e$ at all times.
%
%

\subsection{Electromagnetic spectra}

Using the background evolution of the universe and the time dependence of $I(a)$ given in the previous subsection, we compute the power spectra of the generated electromagnetic fields. To formulate them,
we take the Coulomb gauge, giving $A_0 = \partial_i A_i = 0$, and expand the transverse part of $A_i$ with the polarization
vector $\epsilon_i^{(\lambda)}$ and the creation/annihilation operator 
$a^{\dagger(\lambda)}_{\bm{k}}/a^{(\lambda)}_{\bm{k}}$ as
\footnote{The polarization vector $\epsilon_i^{(\lambda)}$ satisfies
$k_i \epsilon_{i}^{(\lambda)}(\hat{\bm{k}})=0$ and
$\sum_{p=1}^{2} {\epsilon}_{i}^{(\lambda)}(\hat{\bm{k}}) {\epsilon}_{j}^{(\lambda)}(-\hat{\bm{k}})
=\delta_{ij} - (\hat{\bm{k}})_{i}(\hat{\bm{k}})_{j}$,
and the creation/annihilation operators satisfy the commutation relation,
$[a^{(\lambda)}_{\bm{p}},a^{\dagger(\sigma)}_{-\bm{q}}]
= (2\pi)^3\delta(\bm{p}+\bm{q})\delta^{\lambda \sigma}$.}
\begin{equation}
 A_i(\eta, \bm{x})
 = 
 \sum_{\lambda=1}^{2} \int \frac{{\rm d}^3 k}{(2\pi)^3}  \,
 {\rm e}^{i \bm{k \cdot x}} \, 
 \left[ \epsilon_{i}^{(\lambda)}(\hat{\bm{k}}) \,
 a_{\bm{k}}^{(\lambda)} \mcA_{k}(\eta) 
  +   {\rm h.c.} \right]
\,,
\label{introduction of creation/annihilation operator}
\end{equation}
where the hat of $\hat{\bm{k}}$ denotes the unit vector, $(\lambda)$ is the
polarization label and $\mcA_k(\eta)$ is the mode function. 
Note that the mode function ${\cal A}_k (\eta)$ carries no polarization index $(\lambda)$ since the production mechanism in this model does not distinguish different polarization states.
The spectra of electric and magnetic fields are then given by
\begin{equation}
\mcP_E (\eta,k) \equiv \frac{k^3 I^2}{\pi^2 a^4} \, |\partial_\eta \mcA_k|^2 \; ,\qquad
\mcP_B (\eta,k) \equiv \frac{k^5 I^2}{\pi^2 a^4} \, |\mcA_k|^2 \; ,
\label{P of EB}
\end{equation}
respectively.
The equation of motion for the mode function is
\begin{equation}
\left[\partial_\eta^2 +k^2-\frac{ \partial_\eta^2 I }{I}\right](I\mcA_k)=0
\,.
\label{EoM of A}
\end{equation}
From eq.~\eqref{simple I}, it reads
\begin{align}
&\left[\partial_\eta^2 +k^2\right](I_i \mcA_k)=0
\,, \qquad (a<a_i )
\label{BD EoM}
\\ 
&\left[\partial_\eta^2 +k^2-\frac{n(n-1)}{\eta^2}\right](I\mcA_k^\inf)=0
\,, \qquad (a_i <a <a_e)
\label{inf EoM}
\\
&\left[\partial_\eta^2 +k^2-\frac{2n(2n+1)}{\eta^2}\right](I\mcA_k^\osc)=0,
\qquad (a_e <a <a_r)
\label{osc EoM}
\end{align}
where the superscripts ``inf" and ``osc" denote quantities during inflation and the inflaton oscillating phase, respectively.
Provided that the initial condition is given by the Bunch-Davies vacuum state, $I\mcA_k(a<a_i)={\rm e}^{-ik\eta}/\sqrt{2k}$, one can solve $I\mcA_k(a>a_i)$ by using the general solutions of above equations and the junction conditions between them. Then substituting the mode function into eq.~\eqref{P of EB}, one obtains
the electromagnetic power spectra.
Here we show only the super-horizon asymptotic forms of the spectra, while their derivation and exact expressions are shown in appendix \ref{Derivation}:
\begin{align}
\mcP_E^\inf(k,\eta)
&\xrightarrow{|k\eta|\ll 1} \frac{2^{2n+1}}{\pi^4}\Gamma^2(n+1/2)
|C_2( -k\eta_i )|^2 H_\inf^4 |k\eta|^{4-2n},
\label{PE inf}
\\
\mcP_B^\inf(k,\eta)
&\xrightarrow{|k\eta|\ll 1} 
\dfrac{2^{2n-1}}{\pi^4}\Gamma^2(n-1/2)
|C_2( -k\eta_i )|^2 H_\inf^4 |k\eta|^{6-2n},
\\
\mcP_E^\osc(k,\eta)
&\xrightarrow{|k\eta|\ll 1} \frac{2^{2n+1}}{\pi^4}\Gamma^2(n+1/2) 
|C_2( -k\eta_i )|^2
H_\inf^4 \left(\frac{k}{a_e H_\inf}\right)^{4-2n}\left(\frac{a}{a_e}\right)^{2n-4},
\label{PE osc}
\\
\mcP_B^\osc(k,\eta)
&\xrightarrow{|k\eta|\ll 1} 
\dfrac{2^{2n+3}}{\pi^4 }\frac{\Gamma^2(n+1/2)}{(4n+1)^2}
|C_2(-k\eta_i)|^2 H_\inf^4
 \left(\frac{k}{a_e H_\inf}\right)^{6-2n}\left(\frac{a}{a_e}\right)^{2n-3},
\label{PB osc}
\end{align}
where $C_2( x )$ and its asymptotic expressions are given by
\begin{align}
&C_2( x ) =  \frac{i\pi}{2\sqrt{2}}\sqrt{ x } \left[ J_{n-1/2}( x ) -i J_{n+1/2}( x )\right].
\label{fullC2}
\\
&C_2( x ) \xrightarrow{x \ll 1} \frac{i\pi}{2\Gamma(n+1/2)}\left|\frac{ x }{2}\right|^n,
\label{C2 large scale}
\\
&C_2( x ) \xrightarrow{x \gg 1} \frac{i\sqrt{\pi}}{2}
\, {\rm e}^{i \left( -x + \frac{n\pi}{2} \right)} \; .
\label{C2 small scale}
\end{align}
We have also introduced $\eta_i \equiv -1/a_i H_\inf$, which is the conformal time when $I$ starts varying. 

\begin{figure}[tbp]
    \hspace{-2mm}
  \includegraphics[width=70mm]{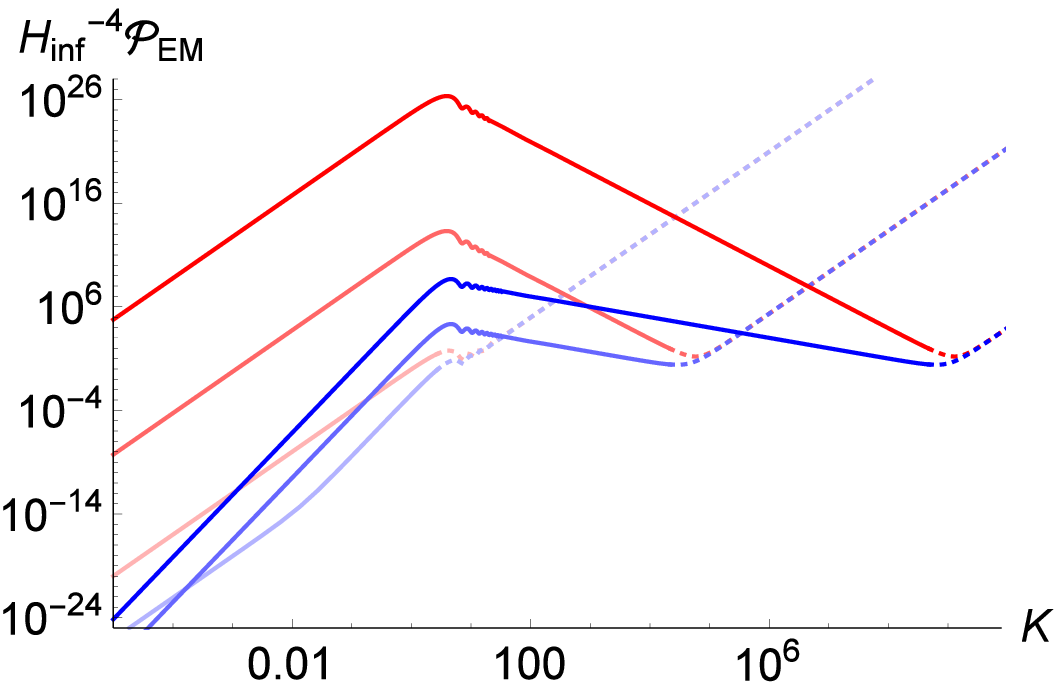}
  \hspace{5mm}
  \includegraphics[width=70mm]{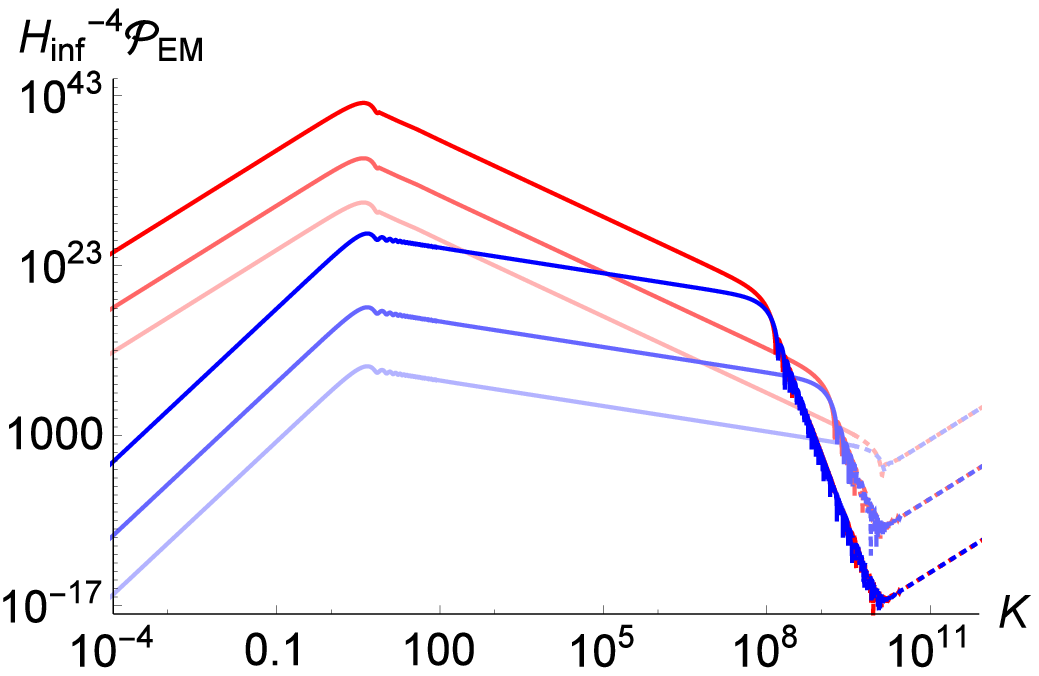}
  \caption
 { {\bf (Left panel)} The electromagnetic spectra during inflation. The horizontal axis denotes $K\equiv |k\eta_i|$, and we set $n=3.5$. $\mcP_E$ (red lines) and $\mcP_B$ (blue lines) normalized by $H_\inf^4$ are shown for $\ln (a/a_i) =1, 10, 20$ from transparent to opaque. The sub-horizon modes are shown as the dotted lines. 
{\bf (Right panel)} The electromagnetic spectra during the inflaton oscillating phase. We set $n=3.5$ and $N_i\equiv \ln(a_e/a_i) \approx22$.
$\mcP_E$ (red lines) and $\mcP_B$ (blue lines) are shown for $\ln (a/a_e) =1,5,10$ from transparent to opaque.
One can see that the sub-horizon modes oscillate and damp, while the super-horizon ones continue to grow. The modes which did not exit the horizon during inflation and thus are unphysical are shown as the dotted line.
}
 \label{EM spectra}
\end{figure}

\begin{figure}[tbp]
  \hspace{-2mm}
  \includegraphics[width=70mm]{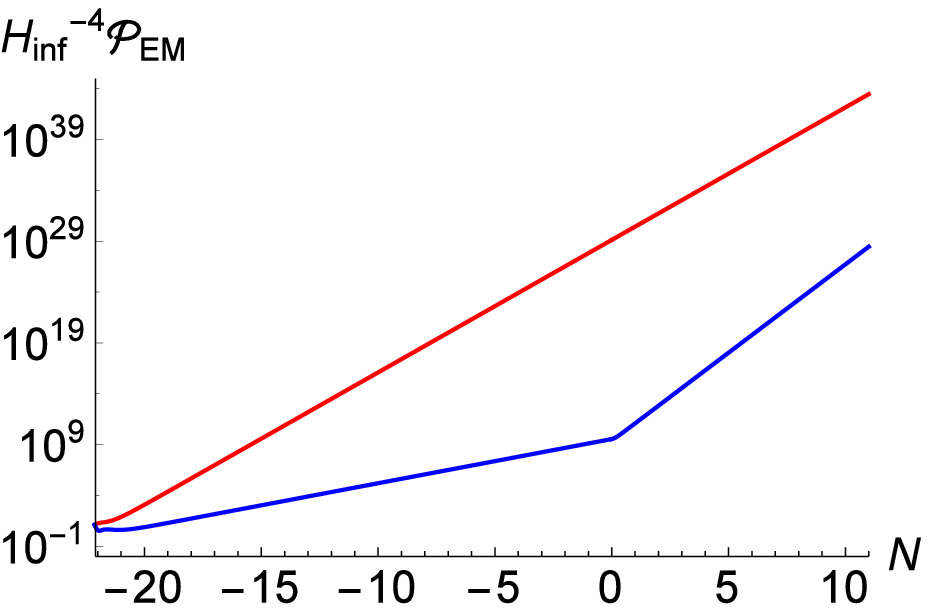}
  \hspace{5mm}
  \includegraphics[width=70mm]{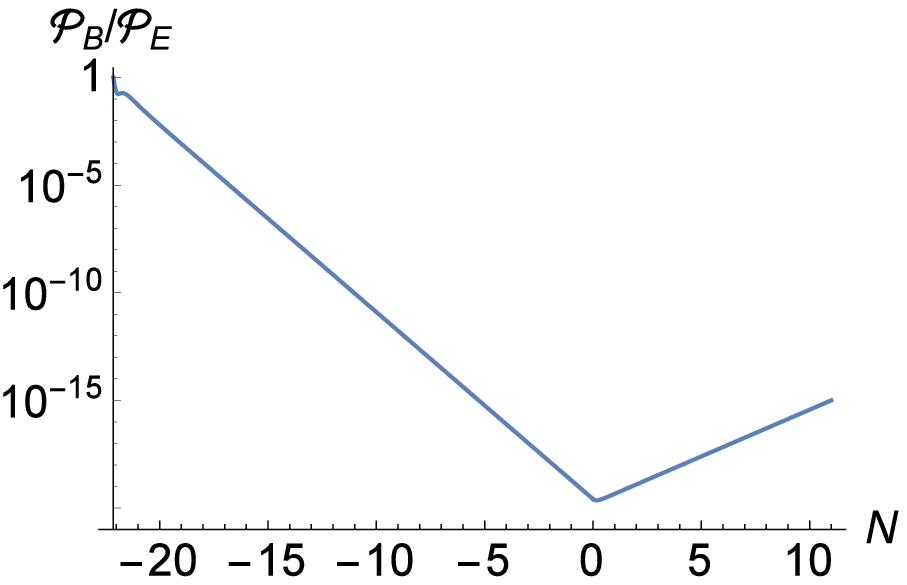}
  \caption
 {{\bf (Left panel)} The time evolution of the electric spectrum $\mcP_E$ (red line) and the magnetic spectrum $\mcP_B$ (blue line). We set $n=3.5$, $N_i\equiv \ln(a_e/a_i)\approx22,$ and $K\equiv|k\eta_i|=4$ which corresponds to the peak scale of $\mcP_E$ and $\mcP_B$.
The horizontal axis denotes the e-folding number $N\equiv\ln(a/a_e)$ and inflation ends at $N=0$. 
{\bf (Right panel)} The ratio between the magnetic and electric power spectrum, $\mcP_B/\mcP_E$. The parameters are the same as the left panel.
The ratio decreases during inflation, but it increases after inflation. }
\label{PEM_evolve}
\end{figure}

In fig.~\ref{EM spectra}, we show the electromagnetic power spectra, $\mcP_E$ and $\mcP_B$, normalized by $H_\inf^4$ for $n=3.5$ during inflation (left panel) and during the inflaton oscillating phase (right panel).
In fig.~\ref{PEM_evolve}, the time evolution of these spectra and their ratio are plotted.%
\footnote{In fig.~\ref{EM spectra} and fig.~\ref{PEM_evolve}, we use the exact solutions in eqs.~\eqref{PEinf Appendix}-\eqref{IPBosc}, instead of the approximated ones.}%
%
%
%
%
The features of the generation of the electromagnetic fields in our scenario are threefold:

\begin{enumerate}

\item[(i)]
{\it Post-inflationary amplification}: In our scenario, it is assumed that $I$ continues to vary after inflation ends. 
This is quite natural if the $\chi$ field that drives $I$ is not the inflaton. As a result, the electromagnetic fields continue to grow even after inflation. Comparing eqs.~\eqref{PE inf}-\eqref{PB osc}, one finds the amplification factors are
\begin{equation}
\frac{\mcP_E^\osc(\eta)}{\mcP_E^\inf(\eta_e)} \simeq \left(\frac{a}{a_e}\right)^{2n-4},
\quad
\frac{\mcP_B^\osc(\eta)}{\mcP_B^\inf(\eta_e)} \simeq \left(\frac{a}{a_e}\right)^{2n-3},
\quad
(|k\eta|\ll1, a\gg a_e).
\label{MF faster}
\end{equation}
They can be substantial amplification for $n>2$. 
Indeed, a massive increase can be seen in  fig.~\ref{EM spectra} and fig.~\ref{PEM_evolve}.
Furthermore, considering that the total energy density decreases in proportional to $a^{-3}$ for $a_e<a<a_r$, and without varying $I$ the magnetic power spectrum would decrease as $a^{-4}$ after inflation, one recognizes that this amplification is very effective for magnetogenesis.
At the same time, however, one may wonder if such a substantial amplification leads to large electric fields which may cause strong backreaction, spoiling the background evolution, or too large curvature perturbation inconsistent with observations. These issues are addressed in the following two points and are discussed in detail in Section \ref{Constraints}.

\item[(ii)]
{\it IR suppression due to the sudden onset of the varying of $I$}: 
As one can see in fig.~\ref{EM spectra}, the spectra are suppressed
on larger scales than $k\sim k_i\equiv |\eta_i|^{-1}$. 
By substituting eq.~\eqref{C2 large scale} into eqs.~\eqref{PE inf}-\eqref{PB osc}, we obtain
\begin{equation}
\mcP_E\propto k^{4},\quad
\mcP_B\propto k^6,\quad (|k\eta_i|\ll1) \; ,
\end{equation}
during both inflation and the inflaton oscillating phase.
In fact, it is advantageous to make the electric power spectrum suppressed on larger scales in order to evade the back reaction and the curvature perturbation problems. 
It is known that in the kinetic coupling model, if one tries to obtain
sufficiently strong magnetic fields on large scales, the electric spectrum should be red-tilted and has a huge amplitude at the IR-cutoff, which corresponds to the mode that crosses horizon at the onset of inflation~\cite{Demozzi:2009fu}.
Then the electric fields cause the problems~\cite{Barnaby:2012xt, Fujita:2013qxa}.%
\footnote{Magnetic fields are always subdominant to electric fields in the scenario where red-tilted magnetic spectra are achieved within the regime free of the strong coupling problem. Thus the backreaction and observational constraints are imposed on the electric field energy.}
In our scenario, however, even if the electric fields have a red-tilted spectrum,
the IR cut-off around $k\sim k_i$ prevents that $\mcP_E$ becomes huge on larger scales~\cite{Ferreira:2013sqa}. In particular, one can expect that
our scenario avoids constraints from the CMB observations if the peak scale $k_i^{-1}$ is much smaller than the CMB scale.

\item[(iii)]
{\it Reduction of the hierarchy between the electric and magnetic fields}:
As one sees in figs.~\ref{EM spectra} and \ref{PEM_evolve}, on super-horizon scales the electric fields are always stronger than the magnetic fields. This is generically true in cases where the gauge mode function is proportional to the power-law of the conformal time, $\mcA_k\propto \eta^s$  on super-horizon scales (in our case, $s=1-2n$ during inflation and $s=1+4n$ during the inflaton oscillating phase). In that case, from the definition of the power spectra eq.~\eqref{P of EB}, one finds
\begin{equation}
\frac{\mcP_B(k,\eta)}{\mcP_E(k,\eta)}=\frac{k^2|\mcA_k|^2}{|\partial_\eta \mcA_k|^2} 
\propto |k \eta|^2
\propto
 \left\{
 \begin{array}{lc}
 a^{-2}, & \quad(\rm inflation)\\
 a, & \quad(\rm oscillation)
 \end{array}\right..
 \label{EB hierarchy}
\end{equation}
Since $|k\eta|\ll 1$ on super-horizon scales, we always have $\mcP_B\ll \mcP_E$.
It should be noted that the hierarchy between $\mcP_E$ and $\mcP_B$ simply depends on how the scale of the mode is bigger than the horizon scale.
Therefore during inflation the hierarchy widens, while it is reduced during the inflaton oscillating phase. This behavior is clearly seen in fig.~\ref{PEM_evolve}. In other words, the magnetic fields grow faster than the electric fields during the inflaton oscillating phase (see eq.~\eqref{MF faster}),
but the opposite is true during inflation.

Since the energy density of the electromagnetic fields is dominated by the electric fields, the constraints coming from the back reaction and the curvature perturbation problem are put on $\mcP_E$. Consequently, stronger constraints are put on $\mcP_B$ because the magnetic fields should be smaller than the electric fields on super-horizon scales by the hierarchical factor, eq.~\eqref{EB hierarchy}.
This is the reason why generated magnetic fields are severely constrained
in conventional inflationary magnetogenesis.
Nevertheless, in our scenario, the hierarchy is reduced by many orders of magnitude during the inflaton oscillating phase (see fig.~\ref{PEM_evolve}). Therefore the constraints on the magnetic fields are substantially relaxed.%
\footnote{By considering very low energy inflation, it is possible to make the hierarchy between $\mcP_E$ and $\mcP_B$ small without the post-inflationary amplification. In this case, however, the curvature perturbation constraint is sensitive to the inflationary dynamics. Hence we do not explore this possibility in this paper.}

\end{enumerate}

\subsection{The strength of the magnetic field at present}
\label{subsec:strength-present}

Let us compute the strength of the produced magnetic field at present and its effective amplitude $B_{\rm eff}$ to compare the prediction of the model with the blazar observation. 
Since the produced magnetic fields evolve adiabatically after $I$ becomes constant at the time of reheating,%
\footnote{If the produced magnetic field is strong enough to satisfy $B_0>10^{-14}\G(\lambda_0/1\rm pc)$, it is possibly processed by the turbulent plasma and its evolution can be modified from the adiabatic evolution~\cite{Durrer:2013pga}. In this paper, however, we focus on the case where the plasma effect is insignificant, and indeed it is the case for the fiducial value in eq.~\eqref{fiducial values}.}
we can obtain the magnetic power spectrum at present by multiplying $\mcP_B^\osc(\eta_r)$ in eq.~\eqref{PB osc} by $a_r^4$, with the scale factor normalized by its present value:
\begin{align}
\mcP_B(\eta_{\rm now})
&=\frac{2^{2n+3}\Gamma^2(n+\frac{1}{2})}{9\pi^4 (4n+1)^2} 
\frac{a_r^4 \rho_\inf^2}{\Mpl^4}
\bigg| C_2 \left( \frac{k}{k_i} \right) \bigg|^2
\, {\rm e}^{2(n-3)N_k + (2n-3)N_r},
\quad (|k\eta_r|\ll 1).
\label{PBnow1}
\end{align}
Here $N_k$ and $N_r$ represent the e-folding number between the horizon crossing of the $k$ mode and the end of inflation, and that of the inflaton oscillating phase, respectively;
\begin{align}
N_r &= \ln \left(\frac{a_r}{a_e}\right) = \frac{1}{3}\ln \left(\frac{H_\inf^2}{H_r^2}\right)
=\frac{1}{3}\ln \left(\frac{\rho_\inf}{\frac{\pi^2}{30}g_* T_r^4}\right)
\notag\\
&\approx 29.5+\frac{4}{3} \ln \left(\frac{\rho_\inf^{1/4}}{10^{10}\GeV}\right)
-\frac{4}{3}\ln \left(\frac{T_r}{1\GeV}\right)-\frac{1}{3}\ln \left(\frac{g_*}{100}\right),
\label{Nr}
\\
N_k &= \ln \left(\frac{a_e H_\inf}{k}\right)=\ln \left(\frac{a_r e^{-N_r}\rho_\inf^{1/2}}{\sqrt{3}\Mpl k}\right)
\notag\\
&\approx 31.4 +\frac{2}{3} \ln \left(\frac{\rho_\inf^{1/4}}{10^{10}\GeV}\right)
+\frac{1}{3}\ln \left(\frac{T_r}{1\GeV}\right)
-\ln \left(\frac{k}{1\Mpc^{-1}}\right) \; ,
\label{Nk}
\end{align}
where  $T_r$ and $g_*$ are the temperature and the number of degree of freedom, respectively, at the time of reheating. We have also used the equation of the entropy conservation,
\begin{equation}
a_r = \left(\frac{g_{\ast s}(T_{\rm CMB})}{g_{\ast s}(T_r)}\right)^{\frac{1}{3}}\frac{T_{\rm CMB}}{T_r}
\approx 8.0 \times10^{-14} \left(\frac{T_r}{1\GeV}\right)^{-1}\left(\frac{g_\ast}{100}\right)^{-1/3},
\label{ar}
\end{equation}
where $T_{\rm CMB}$ is the CMB temperature and $g_{\ast s}$ is the number of degree of freedom for entropy which is assumed to equal to $g_*$ at reheating.

Substituting eqs.~\eqref{Nr}-\eqref{ar} into eq.~\eqref{PBnow1},
we obtain the magnetic power spectrum at present.
One may be tempted to make an immediate comparison with the result of the blazar observations, which actually has been done in some of the literature.
However, it should be stressed that what is measured in the blazar observations is not $\mcP_B$ but $B^2_{\rm eff}$, defined in eq.~\eqref{Beff}.
Therefore we should further substitute the obtained $\mcP_B(\eta_{\rm now})$ into eq.~\eqref{Beff} and compute the effective strength of the magnetic fields. We then obtain
\begin{align}
B_{\rm eff}^2(\eta_{\rm now}) = & 
2 \times 10^2 \, {\rm G}^2 \, \frac{2^{2n} \, {\rm e}^{121.8 (n-2.5)} \, \Gamma^2 \left( n+\frac{1}{2} \right)}{(4n+1)^2} \, H_n (k_i)
\left(\frac{g_*}{100}\right)^{-\frac{2n+1}{3}}
\nonumber\\ & 
\times
\left(\frac{\rho_\inf^{1/4}}{10^{10} \, \GeV}\right)^{4n}
\left(\frac{T_r}{1 \, \GeV}\right)^{-2(n+1)}
\left( \frac{L}{1 \, {\rm Mpc}} \right)^{2n-6} \; ,
\label{Beff2}
\end{align}
where $L \simeq 1 \, {\rm Mpc}$ corresponds to the characteristic length scale for energy losses of charged particles due to inverse Compton scattering, and
\begin{eqnarray}
H_n( k_i ) & \equiv &  \int^{k_{\rm diff}}_0 \frac{{\rm d} k}{k} F(kL)
\bigg| C_2 \left( \frac{k}{k_i} \right) \bigg|^2 (kL)^{2(3-n)} \; ,
\label{Hs def}\\
& \simeq & 
(k_i L)^{5-2n} \exp \left( 8.404 - 2.226 n - 0.1947 n^2 \right),
\label{Hn_fitted}
\end{eqnarray}
where $C_2$ is defined in eq.~\eqref{fullC2}. 
The last approximate expression \eqref{Hn_fitted} which is obtained by fitting the numerical result of \eqref{Hs def} is available in the case with $k_i L\gtrsim1$ and $3 \le n \le 8$. This fit is quite good with error $\sim 1 \, \%$ in this case but not particularly so for $k_i L \ll 1$ or $2.5 \le n \le 3$. Therefore the exponential factor in \eqref{Hn_fitted} that depends only on $n$ is mainly for an illustrative purpose, and we use the exact calculation \eqref{Hs def} for later analyses; however it is worth noting that the $k_i$ dependence, $H_n \propto k_i^{5-2n}$, is quite accurate for $k_i L \gtrsim 1$ even in $2.5 \le n \le 3$.
%
\begin{figure}[tbp]
    \hspace{-2mm}
  \includegraphics[width=70mm]{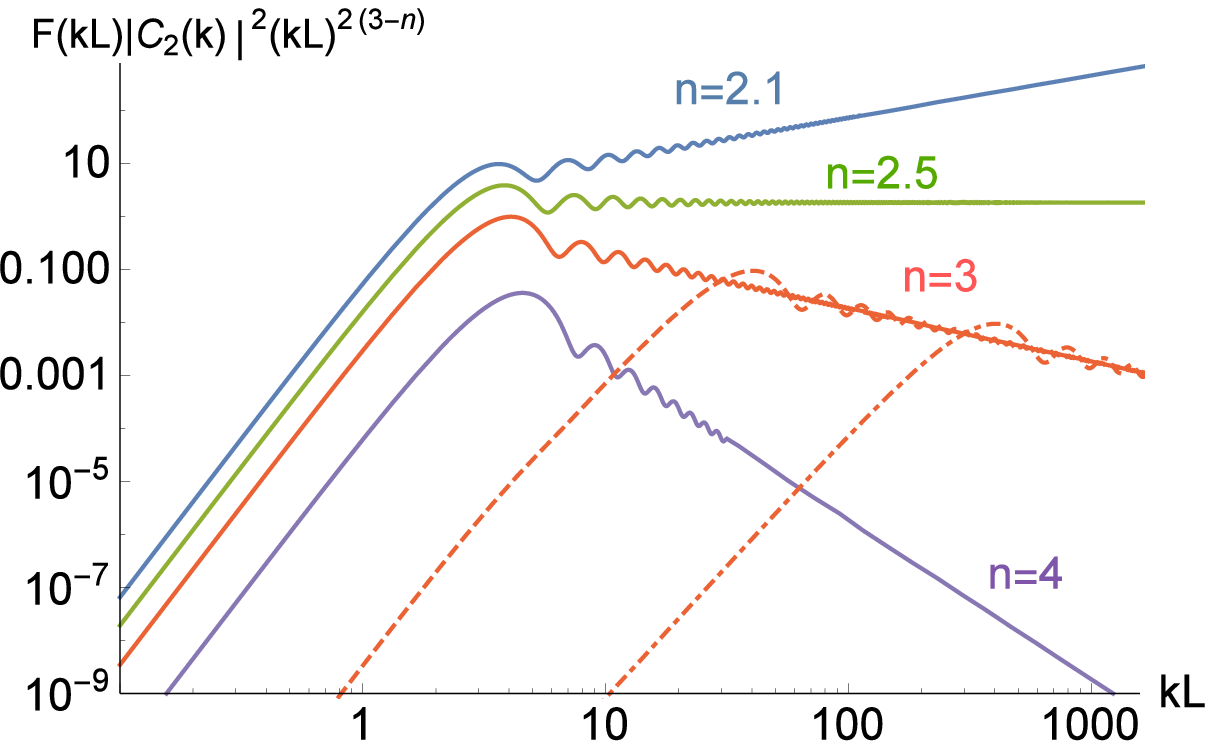}
  \hspace{5mm}
  \includegraphics[width=70mm]{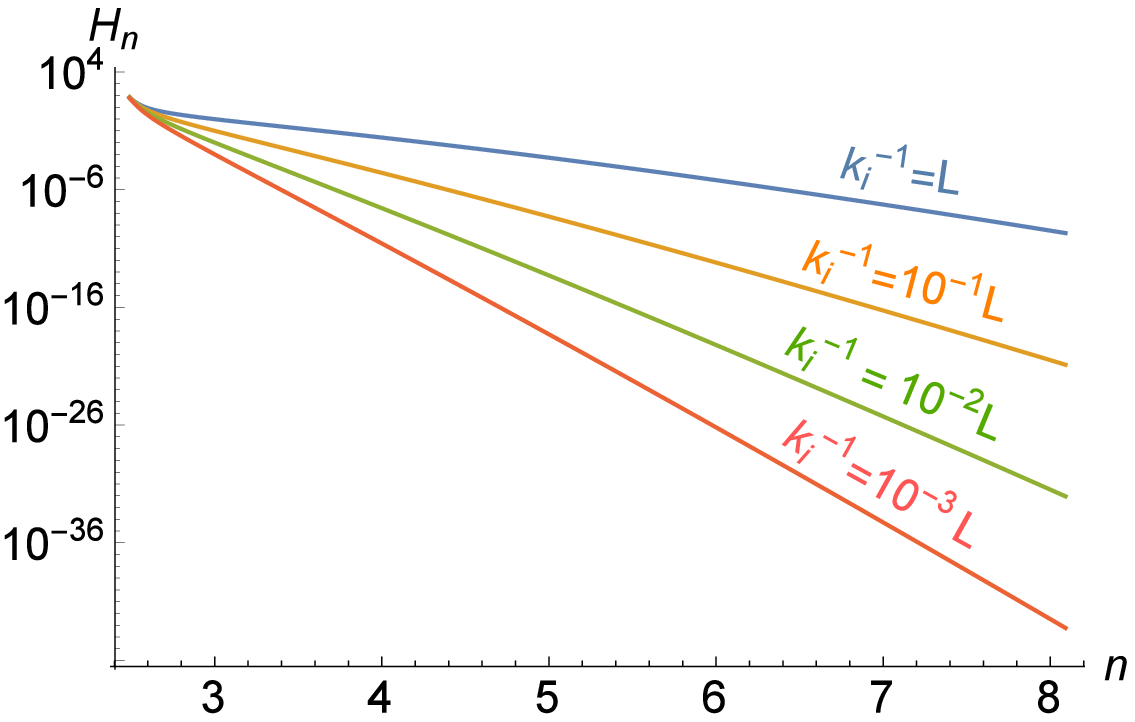}
  \caption
 {{\bf (Left panel)} The integrand of $H_n$ defined in eq.~\eqref{Hs def}. 
 Solid lines are for $k_i^{-1}=L = 1\Mpc$ and $n=2.1, 2.5, 3$ and $4$ from top to bottom. The dashed and dotdashed lines are for $k_i L = 10, 100$, respectively, with $n=3$. One can see that the main contribution to $B_{\rm eff}^2$ comes from $k\sim k_i$ for $n>2.5$.
However, for $n<2.5$, the contribution from  smaller scales is dominant
and $B_{\rm eff}$ depends on the cutoff scale, $k_{\rm diff}$.
 {\bf (Right panel)} The numerically evaluated $H_n$ for $n>2.5$. 
 The lines correspond to $k_i L = 1,10,100$ and $1000$ from top to bottom.
$H_n$ shows the logarithmic divergence for $n=2.5$.
}
 \label{Hs plot}
\end{figure}
%
In fig.~\ref{Hs plot}, we show numerically evaluated $H_n$
and its integrand. It can be seen that for $n<5/2$ the contribution from small scales is dominant and hence $H_n$ depends on the small scale cutoff $k_{\rm diff}$. This is simply because the magnetic power spectrum is blue-tilted for $n<3$ (see eq.~\eqref{PB osc}) and $F(kL) \propto (kL)^{-1}$ for $kL \gtrsim 1$ (see eq.~\eqref{Fz-def}).
Therefore we concentrate on cases with $n>2.5$ henceforth.

\section{Constraints}
\label{Constraints}

We have obtained eq.~\eqref{Beff2} as the magnetic field strength effective for blazar observations in our scenario of the model \eqref{Model Action}. This result should be taken under computational and observational consistencies. In our calculations in the previous sections, we have assumed that the energy density of the produced electromagnetic field does not alter the background evolution to a significant level. We thus have to impose this condition with the obtained result. Moreover, the produced field inevitably contributes to the fluctuations of the total energy density and therefore to the curvature perturbation $\zeta$. Since the electromagnetic spectra are strongly scale-dependent in almost all cases (see fig.~\ref{EM spectra}), the electromagnetically induced $\zeta$ must be subdominant to the standard quasi-scale-invariant curvature perturbation originated from vacuum fluctuations, to be consistent with the CMB observations. One last thing to be taken care of is the effect of charged particles during the reheating process. Some charged particles are produced even before the completion of reheating, and once they are present, they may potentially wash away the electric fields and consequently prevent the evolution of magnetic fields. This effect must be negligible for successful magnetogenesis. We carefully evaluate these three issues one by one in the following subsections.
The final results for present effective magnetic fields with all these constraints imposed are given in Section \ref{Results}.

\subsection{Backreaction problem}

In this subsection, we evaluate the energy density of the produced electromagnetic fields and derive the constraint from the backreaction to the total energy density.
As we discuss in the previous section, the magnetic fields are negligible compared to the electric counterpart, and thus it suffices to focus on the electric fields for the current consideration.
During the inflaton oscillating phase, since the total energy density behaves as $\rho_\tot \simeq \rho_\phi\propto a^{-3}$, eq.~\eqref{PE osc} implies
\begin{equation}
\Omega_\em \equiv \frac{\rho_\em}{\rho_\tot} \propto a^{2n-1} \; ,
\qquad (a_e< a< a_r) \; .
\end{equation}
For $n>1/2$, the energy fraction of the electric fields $\Omega_\em$ increases. In this case, $\Omega_\em$ reaches the maximum value at $a=a_r$, and we should evaluate $\Omega_\em(\eta_r)$. 
Since the contribution from the sub-horizon mode is negligible, we can compute the electric energy density $\rho_E$ for $a_e\le a\le a_r$ from eq.~\eqref{PE osc} as
\begin{align}
\rho_\em^\osc(\eta_r) &\simeq \rho_E^\osc(\eta_r) \simeq \frac{1}{2}\int\frac{\dd k}{k} \mcP^\osc_E(\eta_r),
\notag\\
&\simeq 
\dfrac{2^{2n}}{9\pi^4}\Gamma^2(n+1/2) 
\frac{\rho_\inf^2}{\Mpl^4} \exp\left[(2n-4)(N_i +  N_r)\right]
F_n,
\label{rhoemosc}
\end{align}
where we define
\begin{equation}
N_i\equiv \ln \left(\frac{a_e}{a_i}\right),
\quad
F_n \equiv \int_0^{|\eta_i/\eta_r|} \dd K \, |C_2 (K) |^2 K^{3-2n},
\quad
(K\equiv k/k_i).
\label{Fs def}
\end{equation}
For $n>2$, $F_n$ depends only on $n$, because $\mcP_E^\osc$ has its peak at $k\sim k_i$.
We can numerically evaluate the integral in $F_n$ by sending the upper bound to infinity and find a good fitting function with error $< 1 \, \%$ within the domain $2<n<10$ as
\begin{equation}
F_n \simeq \exp\left( 4.944 -1.461 n - 0.3430 n^2 + 0.0085 n^3 \right) \; .
\label{Fn approx}
\end{equation}
Dividing eq.~\eqref{rhoemosc} by $\rho_r \equiv \rho_\tot(\eta_r)$
and using $\rho_\inf/\rho_r =e^{3N_r}$, we obtain
\begin{align}
\Omega_\em(\eta_r)
&\simeq 
\dfrac{2^{2n}}{9\pi^4}\Gamma^2 \! \left( n+ \frac{1}{2} \right) \,
\frac{\rho_\inf}{\Mpl^4} \exp\left[(2n-4)N_i + (2n-1) N_r\right]F_n,
\notag\\
&\approx
2.5 \times 10^{28} \, 2^{2n} \,  {\rm e}^{121.8(n-2.5)} \, \Gamma^2 \! \left( n+ \frac{1}{2} \right) \, F_n
\notag\\
&\quad\times
\left(\frac{\rho_\inf^{1/4}}{10^{10}\GeV}\right)^{4n}
 \left(\frac{T_r}{1\GeV}\right)^{-2n}
 \left(\frac{g_*}{100}\right)^{\frac{1-2n}{3}}
 \left(\frac{k_i}{1\Mpc^{-1}}\right)^{2(2-n)}.
\label{Omega em 1}
\end{align}
To avoid the backreaction problem, $\Omega_\em(\eta_r)< 1$ is required. Comparing eqs.~\eqref{Beff2} and \eqref{Omega em 1}, one can observe that, to evade strong backreaction, lowering the inflationary energy scale and raising the reheating temperature are favored; however, this would also result in smaller $B_{\rm eff}$. In particular, a higher $T_r$ decreases $B_{\rm eff}$ more than loosening the backreaction, and therefore lowering $\rho_{\rm inf}$ provides a larger parameter window for successful magnetogenesis avoiding the backreaction problem.

\subsection{Curvature perturbation problem}
\label{Curvature perturbation problem}

In this subsection, we explore the curvature perturbation induced by the  production process of the electromagnetic fields, which we call $\zeta_\em$. Considering the curvature perturbation observed in CMB experiments $\zeta_\obs$, the additional contribution to the curvature power spectrum from $\zeta_\em$ must satisfy, $\mcP_\zeta^\em(k_{\rm CMB}) < \mcP_\zeta^\obs(k_{\rm CMB}),$ where
$k_{\rm CMB}$ denotes the CMB scales,
since $\zeta_{\rm em}$ has a strongly scale-dependent spectrum.
This inequality gives a constraint on our magnetogenesis scenario. 
Note that we do not specify the origin of $\zeta_\obs$ and use the observational result $\mcP_\zeta^\obs\approx 2.2 \times 10^{-9}$ in this paper.

On a flat slice (uniform-curvature hypersurface),
the curvature perturbation is given by
\begin{equation}
\zeta = -H\frac{\delta\rho}{\dot{\rho}},
\label{zeta eq1}
\end{equation}
where $\delta\rho$ is density perturbation on the flat slice. 
Here, it is important to notice that the perturbation of the energy density induced by the generation/amplification process of the electromagnetic field includes not only that of the electromagnetic fields itself $\delta\rho_\em\equiv \rho_\em -\langle\rho_\em\rangle$, but also the perturbations of the scalar field energy densities which are sourced by the generated electromagnetic fields. In addition to the direct coupling between the $\chi$ field and the electromagnetic fields, the gravitational interaction couples all fields in our scenario, namely $\phi$, $\chi$ and $A_\mu$.
To properly evaluate the curvature perturbation induced by the produced electromagnetic field, therefore, one must take into account these couplings, solve the equations of motion for the scalar fields, and obtain their energy density perturbations, as well as the direct contribution $\delta\rho_{\rm em}$.

The leading contributions to the scalar perturbations $\delta\phi$ and $\delta\chi$ from the produced electromagnetic field are threefold: (i) the inverse-decay of $A_\mu$ to $\delta\chi$ through the direct coupling $I^2(\chi) F^2$, (ii) $A_\mu$ gravitationally sourcing $\delta\phi$ through the trace of the energy-momentum tensor, and (iii) the gravitational mass mixing of $\delta\phi$ with the sourced $\delta\chi$. These processes can be depicted schematically as (i) $A_\mu + A_\mu \xrightarrow{\rm direct} \delta\chi$, (ii) $A_\mu + A_\mu \xrightarrow{\rm grav.} \delta\phi$, and (iii) $A_\mu + A_\mu \xrightarrow{\rm direct} \delta\chi \xrightarrow{\rm grav.} \delta\phi$, and they and $\delta\rho_{\rm em}$ give contributions of the same order.%
\footnote{The gravitational sourcing to $\delta\chi$, i.e. $A_\mu + A_\mu \xrightarrow{\rm grav.} \delta\chi$ and $A_\mu + A_\mu \xrightarrow{\rm grav.} \delta\phi \xrightarrow{\rm grav.} \delta\chi$, is parametrically smaller and therefore negligible, simply because gravitational interaction is weak and the energy density of $\chi$ is subdominant to that of inflaton $\phi$.}
We refer interested readers to Appendix \ref{app:zeta-em} for the detailed derivation of ${\cal P}_\zeta^{\rm em}$ taking all these effects into account, and only report the final result here. The total energy density perturbation is the sum of all the energy contents,
\begin{equation}
\delta\rho_{\rm tot} = \delta\rho_\phi + \delta\rho_\chi + \delta\rho_{\rm em} \; ,
\end{equation}
and we define the power spectrum of the curvature perturbation in the standard way:
\begin{equation}
\mcP_\zeta (k) \, \frac{2\pi^2}{k^3}  \, (2\pi)^3 \delta^{(3)} (\bm{k}+\bm{k}')
= \left\langle \hat\zeta(\bm{k}) \, \hat\zeta(\bm{k}')\right\rangle
= \frac{H^2}{\dot\rho^2} \left\langle \delta\hat\rho_{\rm tot} (\bm{k}) \, \delta\hat\rho_{\rm tot} (\bm{k}')\right\rangle \; ,
\label{Pzetaem}
\end{equation}
where hat denotes an operator in the Fourier space. Derived in Appendix \ref{app:zeta-em}, the part of $\mcP_\zeta$ sourced directly and indirectly by the produced electromagnetic field, evaluated at the time of reheating, is given in eq.~\eqref{powerzeta-result},
\begin{eqnarray}
{\cal P}_\zeta^\em \big\vert_{t=t_r} &\simeq& 
\frac{\rho_{\rm inf}^2}{M_{\rm Pl}^8} \,
\frac{2^{4n} \, \Gamma^4\left( n + \frac{1}{2} \right)}{3^6 \pi^8} \,
\mathcal{G}_n
\left( \frac{a_e}{a_i} \right)^{4n-5}
\left( \frac{a_r}{a_e} \right)^{4n-2} 
\left( \frac{k}{a_e H_{\rm inf}} \right)^3  \; ,
\label{induced Pzeta}
\end{eqnarray}
where the background time evolution $H_r^2 / H_{\rm inf}^2 = \rho_r / \rho_e \propto (a_e / a_r)^3$ is used, and
\begin{eqnarray}
&& \mathcal{G}_n \equiv
\gamma_n^2 G_n^{(1)}
+ \frac{\pi^4 \lambda_n^2 G_n^{(2)}}{60 \left( 4n+1 \right)^2}
+ \frac{\pi^4 \gamma_n \lambda_n G_n^{(3)}}{12 \left( 4n+1 \right)} \; ,
\\
&& \gamma_n \equiv \frac{8n^2+61n-100}{4(n-2)(2n-1)(4n-5)} \; , \quad
\lambda_n \equiv \frac{3(8n - 7)}{8(2n-1)} \; ,
\nonumber\\ &&
G_n^{(1)} \simeq \exp\left( 5.27 - 2.34 n - 0.821 n^2 + 0.0240 n^3 \right) \; ,
\nonumber\\ && 
G_n^{(2)} \simeq \exp\left( 5.86 - 2.34 n - 0.820 n^2 + 0.0240 n^3 \right) \; ,
\nonumber\\ &&
G_n^{(3)} \simeq \exp\left( 3.46 - 2.33 n - 0.821 n^2 + 0.0241 n^3 \right) \; .
\end{eqnarray}
Note that this expression is valid for $k \ll k_i$, which is relevant for $k \sim k_{\rm CMB}$ and $k_i^{-1} \lesssim 1 \, {\rm Mpc}$.
With eqs.~\eqref{Nr}-\eqref{ar}, eq.~\eqref{induced Pzeta} reads
\begin{align}
\mcP_\zeta^\em \vert_{t=t_r} &\approx
8.7 \times 10^{51} \, 2^{4n} \, \Gamma^4\!\left( n + \frac{1}{2} \right)
{\rm e}^{243.6(n-2.5)} \, 
{\cal G}_n
\notag\\
&\times 
\left(\frac{\rho_\inf^{1/4}}{10^{10}\GeV}\right)^{8n}
 \left(\frac{T_r}{1\GeV}\right)^{-4n}
 \left(\frac{g_*}{100}\right)^{\frac{2(1-2n)}{3}}
 \left(\frac{k_i}{1\Mpc^{-1}}\right)^{5-4n}
  \left(\frac{k}{0.05\Mpc^{-1}}\right)^{3}
\label{Pzetaem-result}
\end{align}
After reheating completes, the electric fields quickly vanish due to the high electric conductivity, and $\zeta_\em$ freezes out~\cite{Martin:2007ue}.
Thus the requirement from the CMB observation is
\begin{equation}
\mcP_\zeta^\em (k_{\rm CMB},\eta_r) < \mcP_\zeta^\obs \approx 2.2\times 10^{-9},
\end{equation}
and this puts a constraint on the strength of the produced magnetic fields.

It should be noted that one can compute higher-order correlation functions of $\zeta_\em$, and they might potentially provide further constraints on the strength of generated electromagnetic fields.
Nevertheless, the curvature perturbation that is sourced by the electromagnetic field is strongly scale-dependent, and
the shape of the bispectrum is very different from the local type or other shapes analyzed in observational papers, such as the ones by the Planck collaboration.
Therefore, the existing bounds on non-Gaussianity are not directly applicable to the present case, and a dedicated work is necessary to obtain a constraint.
Thus we concentrate on the power spectrum in this paper for a concrete analysis, and we would like to come back to this issue in future studies.

\subsection{The interaction with charged particles}

In the previous section, we have solved the equation of motion for the gauge field by ignoring interactions with charged particles. During the inflaton oscillating phase, however, charged particles can be produced by the decay of the inflaton.
If such an interaction is non-negligible, the dynamics of the electromagnetic fields may be significantly altered~\cite{Martin:2007ue,Bassett:2000aw}. Therefore in this subsection, we investigate the condition to safely neglect the interaction which should be satisfied for the consistency of our calculation.
We basically follow the argument in ref.~\cite{Fujita:2015iga},
and assume that the inflaton decay rate $\Gamma_\phi$ is constant. Then the energy density of the decay product is given by $\rho_{\rm rad} =2\Gamma_\phi \rho_\phi/5H$ during the inflaton oscillating phase~\cite{Kolb:1990vq}.

The interaction between photon and charged particles can be ignored if their interaction rate $\Gamma_{\rm int}$ is smaller than the Hubble expansion rate $H$, i.e.  $\Gamma_{\rm int}<H$. One can estimate the interaction rate
as $\Gamma_{\rm int}=n_{\rm c} \sigma_{\rm int} v,$ where $n_{\rm c} \simeq \rho_{\rm rad}/m_\phi$ is the number density of the charged particles,
$\sigma_{\rm int}\simeq \alpha_{\rm eff}^2/m_\phi^2$ is their interaction cross section, and $v\approx 1$ is their velocity.
Here we have introduced the inflaton mass $m_\phi$ and the effective fine structure constant $\alpha_{\rm eff}\equiv \alpha/I^2$ which is rescaled by $I^2$ because the kinetic term of photon is modified~\cite{Demozzi:2009fu}.
Then one can recast the condition $\Gamma_{\rm int}<H$ as the lower bound on $m_\phi$;
\begin{equation}
m_\phi \gtrsim 10^5\GeV\times I^{-\frac{4}{3}}
\left(\frac{\alpha}{0.01}\right)^{2/3}
\left(\frac{T_r}{1\GeV}\right)^{2/3},
\label{mass bound}
\end{equation}
where we have used $T_r \simeq \sqrt{\Gamma_\phi \Mpl}$.
Here we assume the charged particles dominate an $\mathcal{O}(1)$ fraction of $\rho_{\rm rad}$, while the condition eq.~\eqref{mass bound} can be further relaxed if it is not the case. For example, it is possible that the inflaton does not decay into any charged particles and they are only secondarily produced from other decay products~\cite{Fujita:2015iga}.
It is also interesting to note that if the inflaton decays only through dimension five operators, the decay rate is naturally expected to be suppressed by the Planck mass, $\Gamma_\phi \simeq m_\phi^3/\Mpl^2$, and the reheating temperature is given by
\begin{equation}
T_r \simeq \frac{m_\phi^{3/2}}{\Mpl^{1/2}} \approx 1\GeV
\left(\frac{m_\phi}{10^6\GeV}\right)^{3/2}.
\end{equation}
In this case, the condition eq.~\eqref{mass bound} is always satisfied.

\section{Results}
\label{Results}

In this section, we obtain the range of the magnetic field strength in our scenario which evades the backreaction and the curvature perturbation problems.

\begin{figure}[tbp]
    \hspace{-2mm}
  \includegraphics[width=70mm]{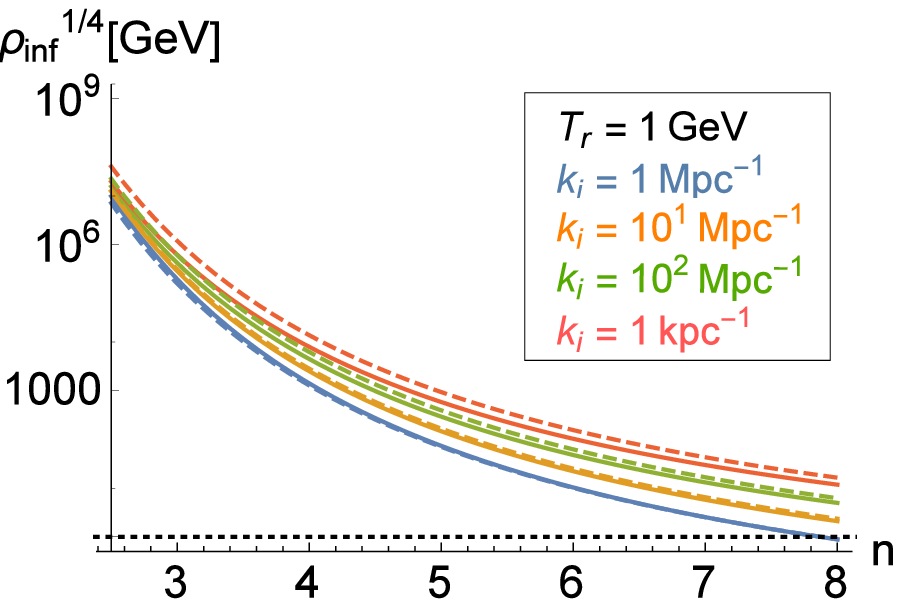}
  \hspace{5mm}
  \includegraphics[width=70mm]{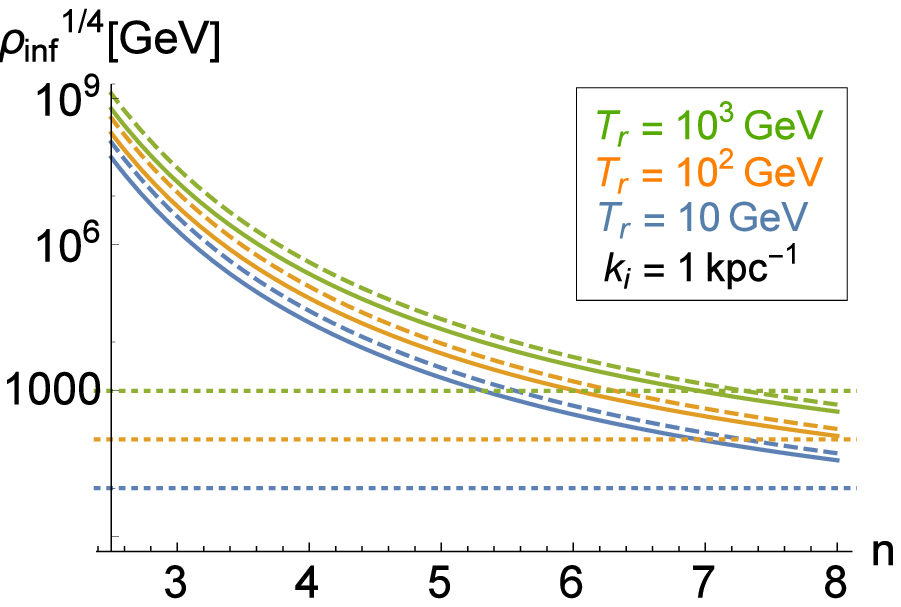}
  \caption
 {{\bf(Left panel)} The inflation energy scale $\rho_\inf^{1/4}$ 
 is shown for $\Omega_\em(\eta_r)=1$ (solid lines) and for $\mcP_\zeta^\em=\mcP_\zeta^\obs$ (dashed lines). 
We fix $k_i= 1$ (blue), $10$ (orange), $10^{2}$ (green) and $ 10^{3} \, \Mpc^{-1}$ (red).
These lines are the upper bounds on the inflationary energy scale, but 
the value of $\rho_{\rm inf}^{1/4}$ depends on the given values of $\Omega_{\rm em}$ and ${\cal P}_\zeta^{\rm em}$ only weakly,
$\rho_\inf^{1/4}\propto \Omega_\em^{1/4n} \propto (\mcP_\zeta^\em)^{1/8n}$.
Since we set $T_r = 1 \, \GeV$, $\rho_\inf^{1/4}$ should be larger than $1 \, \GeV$, shown as the black dashed line. 
{\bf (Right panel)} The peak scale of the electromagnetic field is pushed up to $k_i = 1{\rm kpc}^{-1}$. Again, the solid lines denote $\Omega_\em(\eta_r)=1$ and the dashed lines denote $\mcP_\zeta^\em=\mcP_\zeta^\obs$. The colors represents different reheating temperature, $T_r=10^3$ (green), $10^2$ (orange), and $10 \, \GeV$ (blue). The dotted lines show these $T_r$, setting the lower bounds on $\rho_\inf^{1/4}$. 
}
 \label{rho plot}
\end{figure}

First, we solve eqs.~\eqref{Omega em 1} and \eqref{Pzetaem-result} in terms of $\rho_\inf$ by fixing the parameters, $T_r, g_*$ and $k_i$.
In fig.~\ref{rho plot}, we plot the obtained $\rho_\inf$ for the parameters that we use later.
%
%
Next, substituting the obtained $\rho_\inf$ into eq.~\eqref{Beff2}, we find $B_{\rm eff}$ written in terms of $\Omega_\em(\eta_r)$ and $\mcP_\zeta^\em$ as
\begin{align}
B_{\rm eff}(\eta_{\rm now}) 
\approx& 10^{-13}\G\, 
\frac{\sqrt{H_n/F_n}}{4n+1} \Omega_\em^{1/2}(\eta_r)
 \left(\frac{T_r}{1\GeV}\right)^{-1}
 \left(\frac{g_*}{100}\right)^{-\frac{1}{3}}
 \left(\frac{k_i}{1\Mpc^{-1}}\right)^{n-2},
\label{max B BR}
\\
B_{\rm eff}(\eta_{\rm now}) 
\approx& 10^{-14}\G\, 
\frac{1}{4n+1} \frac{H_n^{1/2}}{\mathcal{G}_n^{1/4}}\left(\frac{\mcP_\zeta^\em}{2.2\times10^{-9}}\right)^{\frac{1}{4}}
\notag\\
&\times \left(\frac{T_r}{1\GeV}\right)^{-1}
 \left(\frac{g_*}{100}\right)^{-\frac{1}{3}}
 \left(\frac{k_i}{1\Mpc^{-1}}\right)^{n-\frac{5}{4}} \left(\frac{k_{\rm CMB}}{0.05\Mpc^{-1}}\right)^{-\frac{3}{4}},
\label{max B CP}
\end{align}
where we have set $L=1 \, \Mpc$.
Note that $H_n$ depends on $k_i$, as we discuss below eq.~\eqref{Hs def}.
To avoid the backreaction and the curvature perturbation problem, it is required that $\Omega_\em(\eta_r)<1$ and $\mcP_\zeta^\em<\mcP_\zeta^\obs$, and these conditions lead to the upper bounds on $B_{\rm eff}$. In fig.~\ref{T=1,k=1},
we show the bounds on $B_{\rm eff}$ for $T_r=1 \, \GeV, g_*=100$ and $k_i=1 \, \Mpc^{-1}$.
%
\begin{figure}[tbp]
  \begin{center}
  \includegraphics[width=90mm]{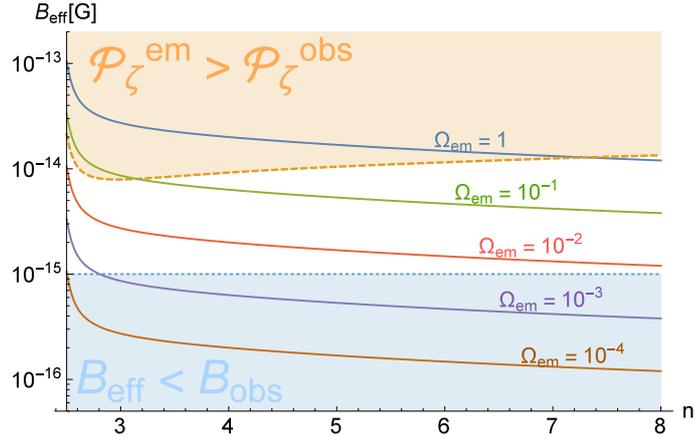}
  \end{center}
  \caption
 { The effective strength of the magnetic field predicted by our model.
 In this figure, we fix the parameters as $T_r=1\GeV, k_i=1\Mpc^{-1}$ and $g_*=100$.  The solid lines represent the cases with $\Omega_\em(\eta_r)=1, 10^{-1}, 10^{-2}, 10^{-3}$ and $10^{-4}$ from top to bottom. The orange shaded region is excluded by the curvature perturbation problem. On the orange dashed line,
the curvature perturbation induced by the electric fields is as large as
the observed one, $\mcP_\zeta^\em = 2.2\times 10^{-9}$. 
The blue dotted line shows the lower bound inferred by blazar observations, $B_{\rm eff} \gtrsim 10^{-15}\G$. The viable region in which sufficient magnetic fields can be generated without the backreaction and the curvature perturbation problem exists for $B_{\rm eff}\lesssim10^{-14}\G$.
}
\label{T=1,k=1}
\end{figure}
%
There exists the viable region where the sufficiently strong magnetic fields, $B_{\rm eff} (\eta_{\rm now})>10^{-15}\G$, are generated without the backreaction or the curvature perturbation problem. The inflationary energy scale corresponding to a given $n$ and $\Omega_\em(\eta_r)$ can be derived from eq.~\eqref{Omega em 1} (or found in fig.~\ref{rho plot}). 
As an illustrative example, the following set of parameters and predictions is obtained in our model:
\begin{align}
n&=3, &T_r&=1\GeV,& k_i&=1\Mpc^{-1}
\notag\\ 
\Omega_\em(\eta_r)&= 1.8 \times 10^{-3},&
\rho_\inf^{1/4}&\approx 1.1\times 10^5 \, \GeV, & B_{\rm eff}(\eta_{\rm now})&\approx 10^{-15} \, \G.
\label{fiducial values}
\end{align}
Note that the generated magnetic fields have a scale-invariant spectrum for $k\gtrsim k_i$ in this case of $n=3$.

Now we explore cases with different $T_r$ and $k_i$.
If the reheating temperature $T_r$ increases, 
the hierarchy between the electric and magnetic fields widens,
since $\mcP_B/\mcP_E(\eta_r)\propto |k\eta_r|^2\propto T_r^{-2}$, and thus the constraints get tighter.
In fact, eqs.~\eqref{max B BR} and \eqref{max B CP} indicate that the maximum $B_{\rm eff}$ decreases in proportional to $T_r^{-1}$.

\begin{figure}[tbp]
\centering
    \hspace{-2mm}
  \includegraphics[width=70mm]{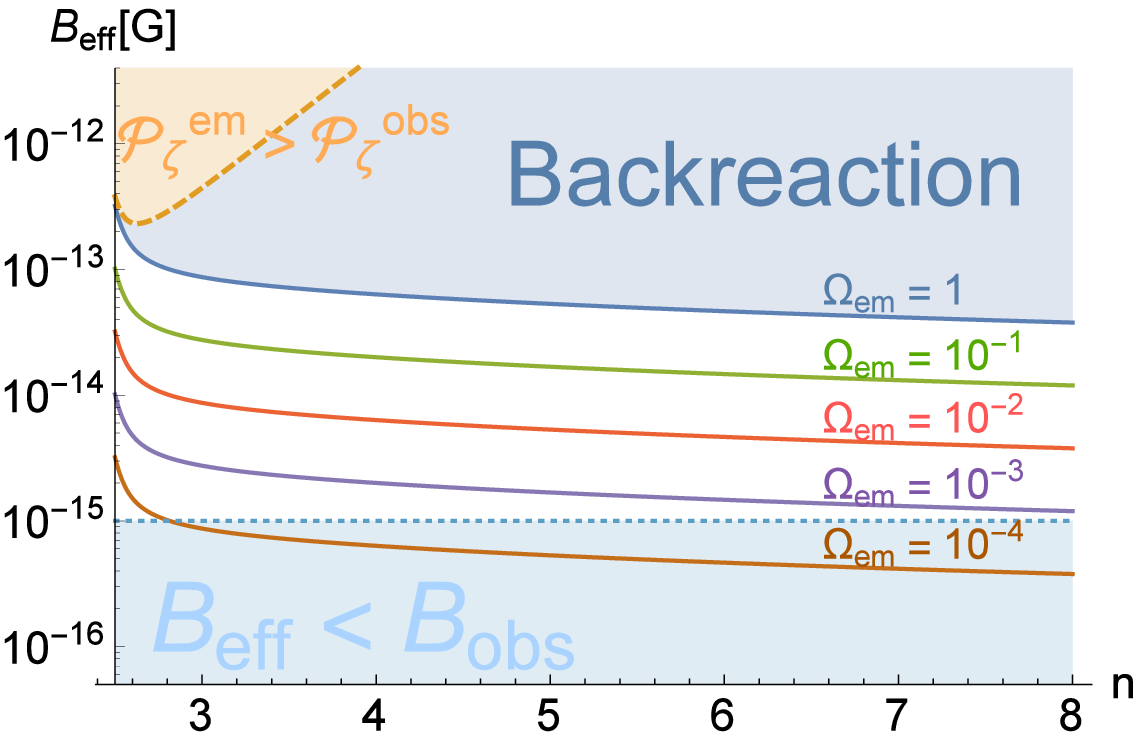}
  \hspace{5mm}
  \includegraphics[width=70mm]{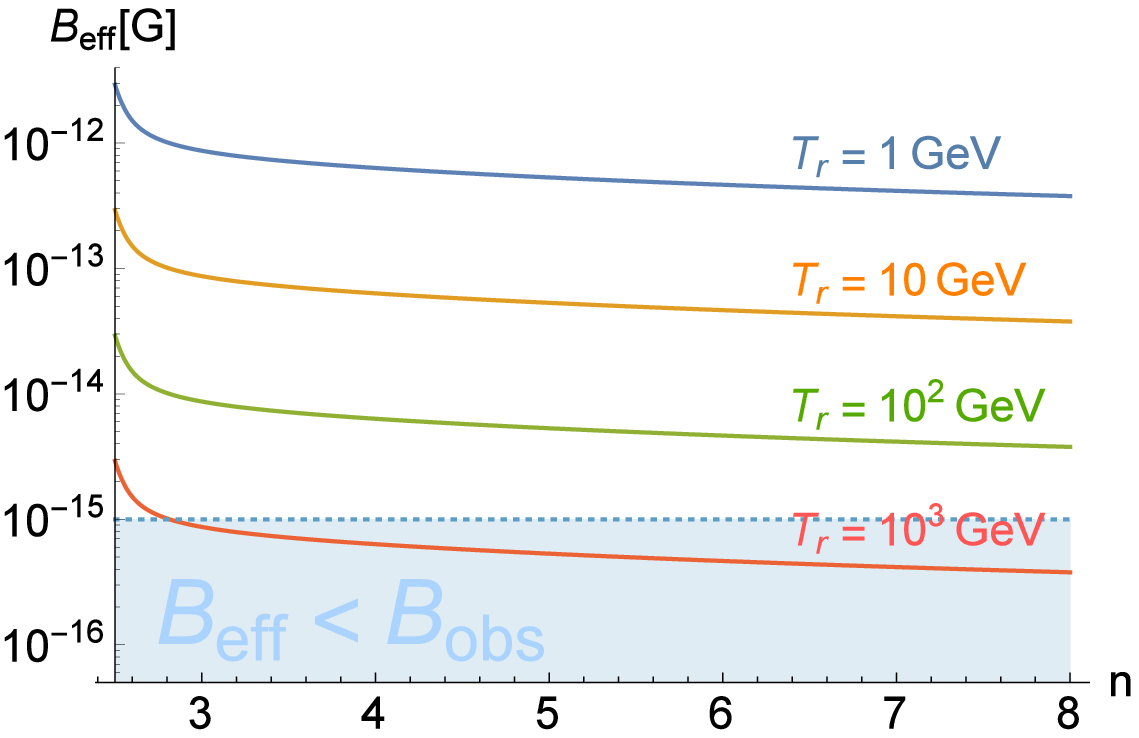}
  \caption
 {{\bf Left panel)} The case with $k_i = 10\Mpc^{-1}$. The other parameters and the plot scheme are the same as fig.~\ref{T=1,k=1}. Since the peak scale of the electromagnetic fields becomes smaller, the hierarchy is more reduced and the constraints are weaker than fig.~\ref{T=1,k=1}. In particular, the constraint from the curvature perturbation becomes irrelevant.
{\bf (Right panel)} The case with $k_i= 1{\rm kpc}^{-1}$. 
The solid lines show $\Omega_\em(\eta_r)=1$ and the reheating temperature
is fixed as $T_r = 1, 10, 10^2$ and $10^3\GeV$ from top to bottom. One can see that the reheating temperature cannot exceed $1$TeV for $k_i \le 1$kpc.
}
 \label{Tk change}
\end{figure}

On the other hand, if one makes $k_i$ larger (i.e.~pushes the peak scale of the electromagnetic fields into a smaller scale), the two constraints become weaker, since the IR cutoff scale goes higher, and thus stronger magnetic fields can be obtained. 
In fig.~\ref{Tk change}, we show the cases with larger $k_i$.
Eqs.~\eqref{max B BR} and \eqref{max B CP} imply that the curvature perturbation constraint becomes irrelevant in comparison with the backreaction constraint for sufficiently large $k_i$. This is because
the characteristic scale of the electromagnetic fields goes away from the CMB scale. Furthermore, since the hierarchy between electric and magnetic fields is reduced, the back reaction problem is also relaxed.
This time, however, a simple scaling argument is difficult because a varying $k_i$ (or $\eta_i$) changes the numerical integration of $H_s$ in $B_{\rm eff}$ (see eq.~\eqref{Hs def} and fig.~\ref{Hs plot}). 
Comparing fig.~\ref{T=1,k=1} and the left panel of fig.~\ref{Tk change}, one can see how the viable region broadens by pushing $k_i$ from $1 \, \Mpc^{-1}$ into $10 \, \Mpc^{-1}$.
In the right panel of fig.~\ref{Tk change}, we show the present values of the effective field strength $B_{\rm eff}(\eta_{\rm now})$ with the backreaction constraint saturated ($\Omega_\em(\eta_r)=1$), with $k_i=1 \, {\rm kpc}^{-1}$ for $T_r=1, 10, 10^2$ and $10^3 \, \GeV$.
%
\begin{figure}[t]
\centering
\includegraphics[width=80mm]{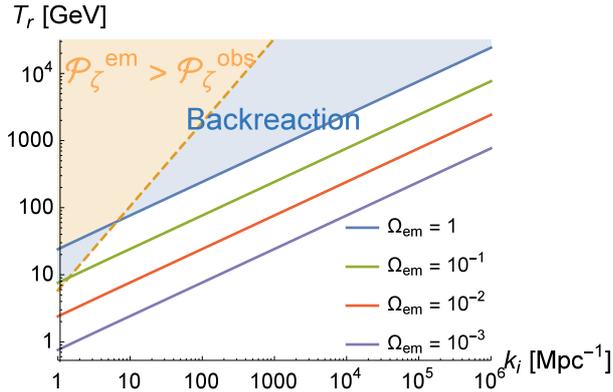}
\caption{The reheating temperature as a function of $k_i$. Here we fix $B_{\rm eff} (\eta_{\rm now}) = 10^{-15} \, {\rm G}$, choose $n=3$, which gives a scale-invariant magnetic spectrum, and show a few cases for different values of $\Omega_{\rm em} (\eta_r)$.  As can be seen from \eqref{Tr approx}, increasing $k_i$ can achieve higher $T_r$ for a given $B_{\rm eff}$ and $\Omega_{\rm em}$. 
}
\label{fig:Tk_ki}
\end{figure}
%

Finally, let us comment on the allowed maximum reheating temperature in this scenario. 
Combining eqs.~\eqref{Hn_fitted}, \eqref{Fn approx} and \eqref{max B BR},
one can derive an approximated expression for the reheating temperature as
\begin{equation}
T_r \simeq 5.6\times 10^2 \, \GeV \, \frac{e^{-0.3825n+0.07415n^2-0.00425n^3}}{4n+1} \,
\Omega_\em^{1/2} 
\left(\frac{k_i}{1\Mpc^{-1}}\right)^{1/2}
\left(\frac{B_{\rm eff}}{10^{15}\G}\right)^{-1}
\left(\frac{g_*}{100}\right)^{-\frac{1}{3}}.
\label{Tr approx}
\end{equation}
It should be noted that this expression is valid for $k_i L\gtrsim1$ and $3\le n \le 8$.
As seen from this expression and fig.~\ref{fig:Tk_ki}, higher reheating temperature can be achieved for larger values of $k_i$. In this figure, we fix $B_{\rm eff} (\eta_{\rm now}) = 10^{-15} \, {\rm G}$ and $n=3$, which leads to a scale-invariant magnetic spectrum with an observed amplitude, and take a few different values of $\Omega_{\rm em}(\eta_r)$. 
Interestingly, given the $k_i$ dependence of $H_n$ as in \eqref{Hn_fitted}, one can see $T_r \propto k_i^{1/2}$, independent of the value of $n$, in the range $k_i L \gtrsim 1$. The constraint from the curvature perturbation quickly becomes irrelevant as $k_i$ goes further away from the CMB scales. Hence the possible range of reheating temperature is not much limited in the model.%
\footnote{
The requirement of $k_i$ for the galactic seed magnetic fields is not clear and $k_i \gg 1 \, {\rm kpc}^{-1}$ is not excluded. Furthermore, the necessity of the primordial seed magnetic field for the galactic magnetic fields is also disputed. If it is not necessary, $k_i$ could be larger.}

\section{Conclusion}
\label{Summary}

We investigate the viability of successful magnetogenesis in the model of the electromagnetic field coupled to a non-inflaton scalar field $\chi$ through its kinetic term in the primordial universe.%
\footnote{Here we say ``electromagnetic field'' to mean a Standard Model $U(1)$ gauge field. In principle, if the production occurs before the electroweak phase transition, one should instead consider the gauge field associated with the $U(1)$ hypercharge. The true electromagnetic field consists partially of this gauge field, and the conversion is trivial. This may modify the strength of the magnetic field only around 10\% and therefore will not change our conclusion.}
The time variation of the kinetic function $I(\chi)$ transfers the energy of $\chi$ into the electromagnetic field and thus leads to the production of photons.
We assume that $I(\chi)$ evolves in time only for a fixed period during inflation and continues until the completion of reheating; after reheating, the electromagnetic field evolves adiabatically. The produced magnetic field is originated from the vacuum fluctuations during inflation, and its scale dependence differs among different modes, which can be classified into four cases: (i) the modes that are always outside horizon from the onset of $I(\chi)$ until reheating, (ii) those that cross horizon after the onset of $I(\chi)$ and stay super-horizon until reheating, (iii) those that both cross and re-enter horizon between the onset of $I(\chi)$ and the time of reheating, and (iv) those that do not exit horizon until the end of inflation and thus never do.

Our aim is to search for a successful scenario for the generation of large-scale magnetic fields to account for the blazar observations, preserving all the computational and observational consistencies, namely respecting the backreaction and the CMB constraints within a weak-coupling regime. 
Although a red-tilted magnetic spectrum is preferred for large-scale fields, this, combined with the requirement to avoid the strong-coupling problem, results in much larger production of electric fields. Hence imposing the constraints on them largely suppresses the amplitude of the corresponding magnetic fields. 
It has been known to be particularly difficult, if not impossible, to generate $\sim 1 \, {\rm Mpc}$ scale magnetic fields solely from inflation, unless inflationary energy scale is extremely low, around the BBN bound.
In view of the blazars, we choose such parameters that the scale for the peak amplitude is of ${\cal O}(1 \, {\rm Mpc})$ or smaller, and then the scales relevant for the CMB observations are much larger, corresponding to the modes (i) in the previous paragraph. In this case, the constraints from the CMB curvature perturbation are relatively loose, while the produced magnetic fields keep evolving after inflation, preventing their dilution against the background energy density during the period of inflaton oscillation.

We compute the effective amplitude of the present magnetic field, imposing the constraints from the CMB fluctuations and from the backreaction to the background dynamics, in this rather optimal scenario. To compute the curvature perturbation induced by the produced electromagnetic field, we include all the relevant contributions, namely those coming from the direct coupling $I^2 F^2$ and from the gravitational interactions, up to the leading order.  We restrict our attention to the two-point correlator of the curvature perturbation and require the sourced part of its power spectrum to be smaller than the observed quasi scale-invariant value ${\cal P}_\zeta^{\rm obs} \cong 2.2 \times 10^{-9}$, since the sourced mode is strongly scale-dependent. The shape of the induced non-Gaussianity is different from that of the templates used in the CMB analysis, and the existing bounds on non-Gaussianity in the Planck papers are not directly applicable. Therefore a dedicated data analysis 
 would be
  necessary to provide a constraint on our mechanism of magnetogenesis from higher-order correlation functions, which is beyond the scope of this paper.

We find a viable parameter space for the generation of magnetic fields with amplitudes sufficient to account for the spectrum of the $\gamma$ rays from distant blazars. This is, to our knowledge, the first example of successful large-scale magnetic fields of primordial origin in the $I^2 F^2$ model with inflationary energy scales away from the BBN, respecting all the observationally relevant constraints consistently in the weak-coupling regime. The constraint from the curvature perturbation places the strongest bounds if the peak scale of the produced magnetic field is ${\cal O}(1 \, {\rm Mpc})$. The smaller the peak scale is, however, the looser both the backreaction and the CMB constraints are, as the power on larger scales is more suppressed. 
We also verify that the conductivity induced by the charged particles that may be present even before the completion of reheating does not prevent the evolution of the magnetic fields, since the effective electromagnetic coupling $e / I$ is much smaller than unity before this time.
Our results also infer that the reheating temperature for successful magnetogenesis has a strong relationship with the peak scale of the magnetic field. If one allows a small correlation length of the magnetic field, still compatible with the observed amplitude, then a rather large range of reheating temperature can be realized.

While our scenario succeeds to generate magnetic fields large enough for blazar observations, it still lacks a concrete model. We have assumed a simple time dependence of background $I(a)$, but we have not specified the functional form of $I(\chi)$ or of the potential $U(\chi)$ to realize the desired time dependence. This model building would require a further investigation and is beyond the scope of our current goal, which is to provide a successful scenario for primordial magnetogenesis.
In a realistic scenario, it would not be surprising that the time dependence of $I$ changes at the end of inflation, and then the magnetic-field spectrum would have more non-trivial shape. Also the decay time of $\chi$ would not necessarily coincide with that of the inflaton; under some circumstances, $\chi$ might behave as a curvaton, which would be an interesting possibility.
The construction and analysis of a realistic model, as well as potential constraints from higher-order correlations, are important issues that we would like to come back in the future studies.

\acknowledgments

The authors are grateful to Takeshi Kobayashi, Sabino Matarrese and Marco Peloso, Jun'ichi Yokoyama and Shuichiro Yokoyama for useful discussions. This work is supported in part by the World Premier International
Research Center Initiative (WPI Initiative), MEXT, Japan. 
The work of TF has been supported in part by the JSPS Postdoctoral Fellowships for Research Abroad (Grant No. 27-154).

\appendixpage
\appendix

\section{Derivation of the Electromagnetic Power Spectra}
\label{Derivation}

In this appendix, we solve the E.o.M. for the mode function $\mcA_k$, namely
eqs.~\eqref{BD EoM}-\eqref{osc EoM}, and obtain the electromagnetic power spectra, $\mcP_E$ and $\mcP_B$.
First we assume the gauge field is in the Bunch-Davies vacuum state for $a<a_i$,
\begin{equation}
I_i\mcA_k^\BD (a<a_i) =\frac{e^{-ik \left( \eta - \eta_i \right)} }{\sqrt{2k}} \; ,
\label{BD vac}
\end{equation}
where the constant phase is added for convenience.
Solving eqs.~\eqref{inf EoM} and \eqref{osc EoM} one finds the general solutions are given by
\begin{align}
&I\mcA_k^\inf(\eta) =\sqrt{-\eta}\left[ C_1 J_{n-1/2} (-k\eta)+ C_2 Y_{n-1/2} (-k\eta)\right],
\label{inf general solutions}
\\
&I\mcA_k^\osc (\eta) = \sqrt{\eta} \left[ D_1 J_{2n+1/2} (k\eta) + D_2 Y_{2n+1/2}(k\eta) \right],
\label{osc general solutions}
\end{align}
where  $J_\nu(x)$ and $Y_\nu(x)$ are the Bessel function of the first and second kind, respectively. Here $C_1, C_2, D_1$ and $D_2$ are constant and they will be determined by the junction conditions.
By using the junction condition between eqs.~\eqref{BD vac} and \eqref{inf general solutions}  at $\eta=\eta_i$,
\footnote{Since the electric energy density depends on not $\partial_\eta(I\mcA_k)$
but $I\partial_\eta \mcA_k$, the variable which should be used in the junction is not $\partial_\eta (I\mcA_k)$ but $\partial_\eta \mcA_k$ to ensure the continuity of the physical quantity.}
\begin{equation}
\mcA_k^\BD(\eta_i) = \mcA_k^\inf(\eta_i),
\qquad
\partial_\eta\mcA_k^\BD(\eta_i) = \partial_\eta\mcA_k^\inf(\eta_i),
\end{equation}
one finds $C_1$ and $C_2$ are given by 
\begin{align}
C_1 &= - \frac{i\pi}{2\sqrt{2}}\sqrt{-k\eta_i} \left[ Y_{s-1/2}(-k\eta_i) -i Y_{s+1/2}(-k\eta_i)\right],
\label{fullC1}
\\
C_2 &=  \frac{i\pi}{2\sqrt{2}}\sqrt{-k\eta_i} \left[ J_{s-1/2}(-k\eta_i) -i J_{s+1/2}(-k\eta_i)\right].
\end{align}

Next, one can connect eq.~\eqref{inf general solutions} to eq.~\eqref{osc general solutions} by using the junction condition at the end of inflation, $a=a_e$. It should be noted that the conformal time $\eta$ is not continuous there.
Requiring that the scale factor $a$ and Hubble parameter $H$ are continuous,
one finds $\eta$ jumps as
\begin{equation}
\eta_e = -\frac{1}{a_e H_\inf} \quad{\rm (end\ of\ inflation)}
\quad\Longrightarrow\quad
\tilde{\eta}_e= \frac{2}{a_e H_\inf}\quad {\rm (onset\ of\ oscillation)}.
\end{equation}
Thus the junction condition is 
\begin{equation}
\mcA_k^{\inf}(\eta_e) = \mcA_k^{\rm MD}(\tilde{\eta}_e),
\qquad
\partial_\eta\mcA_k^{\inf}(\eta_e) = \partial_\eta\mcA_k^{\rm MD}(\tilde{\eta}_e),
\end{equation}
and one can obtain $D_1$ and $D_2$.
The calculation is straightforward while the full expressions of $D_1$ and $D_2$ are complicated. Since we are interested only in the modes which exit the horizon during inflation, we show their asymptotic form in the limit $|k \eta_e|\ll 1$ 
(we use the full expression to plot Fig.~\ref{EM spectra}); 
\begin{align}
D_1 &\simeq - \frac{2^n}{\pi} \Gamma(2n+1/2)\Gamma(n+1/2) |k\eta_e|^{-3n} C_2,
\label{D1}
\\
D_2 &\simeq  2^{n-2} \frac{3\Gamma(n-1/2)}{\Gamma(2n+3/2)}|k\eta_e|^{1+n}C_2
-\frac{2^{-n}\pi |k\eta_e|^{3n}C_1}{\Gamma(2n+1/2)\Gamma(n+1/2)}.
\label{D2}
\end{align}
The second term in eq.~\eqref{D2} is important only for very large scale modes, $|k\eta_e|(\eta_i/\eta_e)^{2n}\lesssim 1$, and thus we ignore it.

Now we can obtain the electromagnetic power spectra by substituting eqs.~\eqref{inf general solutions} and \eqref{osc general solutions} into eq.~\eqref{P of EB}.
Let us see the results in order.

\subsection{During inflation, before $I$ starts varying: $a<a_i$} 

Substituting the Bunch-Davies vacuum eq.~\eqref{BD vac}, one finds 
\begin{equation}
\mcP_E^{\rm BD} =\mcP_B^{\rm BD}= \frac{H_\inf^4}{2\pi^2} |k\eta|^4.
\end{equation}
In the vacuum, the magnetic and the electric spectrum are identical, and they are blue-tilted in proportional to $k^4$. Even after $I$ starts varying, the electromagnetic spectra on sub-horizon scales do not change. This is because the $k^2$ term dominates the $\partial_\eta^2 I/I$ term in eqs.~\eqref{inf EoM} and \eqref{osc EoM}, and  thus the sub-horizon modes do not feel the time-variation of $I$.
Hence, hereafter we focus on the super-horizon modes.

\subsection{During inflation, after $I$ starts varying: $a_i<a<a_e$} 

Substituting  eq.~\eqref{inf general solutions} into eq.~\eqref{P of EB}, one finds 
\begin{align}
\mcP_E^\inf
&= \frac{H_\inf^4}{\pi^2} |k\eta|^5 \left| C_1 J_{n+1/2}(-k\eta)+C_2 Y_{n+1/2}(-k\eta)\right|^2,
\quad ({\rm exact})\notag\\
&\xrightarrow{|k\eta|\ll 1} \frac{2^{2n+1}}{\pi^4}\Gamma^2(n+1/2)
|C_2|^2 H_\inf^4 |k\eta|^{4-2n}, \quad ({\rm super\ horizon})
\notag\\
&\simeq
\left\{
 \begin{array}{lc}
 \dfrac{2^{2n-1}}{\pi^3}\Gamma^2(n+1/2)H_\inf^4 |k\eta|^{4-2n},
 & \quad(|k\eta_i|\gg 1)\\
 \dfrac{H_\inf^4}{2\pi^2}\left(\dfrac{\eta_i}{\eta}\right)^{2n}|k\eta|^4,  & \quad (|k\eta_i|\ll1)
 \end{array} 
 \right. \quad (C_2\ {\rm approx.}).
 \label{PEinf Appendix}
\end{align}
\begin{align}
&\mcP_B^\inf
= \frac{H_\inf^4}{\pi^2} |k\eta|^5 \left| C_1 J_{n-1/2}(-k\eta)+C_2 Y_{n-1/2}(-k\eta)\right|^2,
\quad ({\rm exact})\notag\\
&\xrightarrow{|k\eta|\ll 1} 
\dfrac{2^{2n-1}}{\pi^4}\Gamma^2(n-1/2)
|C_2|^2 H_\inf^4 |k\eta|^{6-2n}, \quad  ({\rm super\ horizon})
\notag\\
&\simeq
\left\{
 \begin{array}{lc}
 \dfrac{2^{2n-3}}{\pi^3}\Gamma^2(n-1/2)H_\inf^4|k\eta|^{6-2n},  & \quad (|k\eta_i|\gg1)  \\
 \dfrac{H_\inf^4}{2\pi^2(2n-1)^2}\left(\dfrac{\eta_i}{\eta}\right)^{2n}|k\eta|^{6},
 & \quad(|k\eta_i|\ll\ 1)
  \end{array} 
 \right. \quad (C_2\ {\rm approx.}).
\end{align}
In both the equations, the first line shows the exact expression, the second line shows the super-horizon asymptotic formula, and in the third line
the asymptotic form of $C_2$, eqs.~\eqref{C2 large scale} and \eqref{C2 small scale}, are used. Exactly speaking, right after $a=a_i$, $\mcA_k$ is dominated by a constant term for a short interval, $\eta_i \le \eta < \eta_c$ with
$|k\eta_c| \equiv [(2 n -1)\pi^2]^{\frac{-1}{2 n -1}}|k\eta_i|^{\frac{2 n}{2 n -1}}$.
That term gives an additional contribution to $\mcP_B^\inf$. However,
we suppress it in the approximated expressions because it quickly becomes subdominant for $k\sim k_i$ and is not significant to estimate the final amplitude of the magnetic fields.

\subsection{Inflaton oscillating phase with varying $I$: $a_e<a<a_r$}

Substituting  eq.~\eqref{osc general solutions} into eq.~\eqref{P of EB}, one finds 
\begin{align}
\mcP_E^\osc
&= \frac{H_\inf^4}{2^4\pi^2} |k\eta|^5 \left(\frac{\tilde{\eta}_e}{\eta}\right)^{12} \left|D_1 J_{2n-1/2}(k\eta)+D_2 Y_{2n-1/2}(k\eta)\right|^2,
\quad ({\rm exact})\notag\\
&\xrightarrow{|k\eta|\ll 1} \frac{2|D_1|^2}{\pi^2\Gamma^2(2n+1/2)}
 H_\inf^4 |k\eta_e|^{4(n+1)}\left(\frac{\eta}{\tilde{\eta}_e}\right)^{4(n-2)}, \quad ({\rm super\ horizon})
\notag\\
&\simeq
\left\{
 \begin{array}{lc}
 \dfrac{2^{2n-1}}{\pi^3}\Gamma^2(n+1/2)H_\inf^4 |k\eta_e|^{4-2n}\left(\dfrac{\eta}{\tilde{\eta}_e}\right)^{4(n-2)},
 & \quad(|k\eta_i|\gg 1)\\
 \dfrac{H_\inf^4}{2\pi^2}|k\eta_e|^4\left(\dfrac{\eta_i}{\eta_e}\right)^{2n}\left(\dfrac{\eta}{\tilde{\eta}_e}\right)^{4(n-2)},  & \quad (|k\eta_i|\ll1)
 \label{IPEoscC2}
 \end{array} 
 \right. \quad (C_2\ {\rm approx.}).
\end{align}
\begin{align}
& \mcP_B^\osc
= \frac{H_\inf^4}{2^4\pi^2} |k\eta|^5 \left(\frac{\tilde{\eta}_e}{\eta}\right)^{12}  \left| D_1 J_{2n+1/2}(k\eta)+D_2 Y_{2n+1/2}(k\eta)\right|^2,
\quad ({\rm exact})\notag\\
&\xrightarrow{|k\eta|\ll 1} 
\dfrac{2|D_1|^2}{\pi^2\Gamma^2(2n+3/2)}H_\inf^4
|k\eta_e|^{4n+6} \left(\frac{\eta}{\tilde{\eta}_e}\right)^{4n-6} ,  \quad({\rm super\ horizon})
\notag\\
&\simeq
\left\{
 \begin{array}{lc}
 \dfrac{2^{2s+1}\Gamma^2(n+1/2)}{\pi^3(4n+1)^2}H_\inf^4|k\eta_e|^{6-2n}\left(\dfrac{\eta}{\tilde{\eta}_e}\right)^{4n-6},  & \quad (|k\eta_i|\gg1)  \\
 \dfrac{2H_\inf^4}{\pi^2(4n+1)^2}|k\eta_e|^{6}\left(\dfrac{\eta_i}{\eta_e}\right)^{2n} \left(\dfrac{\eta}{\tilde{\eta}_e}\right)^{4n-6},
 & \quad(|k\eta_i|\ll\ 1)
 \end{array} 
 \right. \quad (C_2\ {\rm approx.}).
 \label{IPBosc}
\end{align}
Again we ignore the constant contribution to $\mcA_k^\osc$ on super-horizon scales, which becomes dominant for a short interval while it is not relevant to the final result.
Comparing these spectra, one can explicitly confirm the hierarchical relation
eq.~\eqref{PE osc}, $\mcP_B \sim |k\eta|^2 \mcP_E$ holds
during both inflation and the inflaton oscillating phase.

\section{The calculation of $\zeta_\em$} 
\label{app:zeta-em}

We have discussed the constraints due to the CMB observations in Subsection \ref{Curvature perturbation problem} of the main text, where only the final results are reported.
In this appendix section, we show the detailed calculation of the curvature perturbations induced by the produced photon fields. To this end, we expand the
inflaton ($\phi$) and spectator ($\chi$) fields as
\begin{equation}
\phi = \phi_0 + \delta\phi \; , \quad \chi = \chi_0 + \delta\chi \; ,
\end{equation}
and decompose the scalar part of the metric in the spatially flat gauge,
\begin{equation}
g_{00} = - a^2 \left( 1 + 2 \, \Phi \right) \; , \quad
g_{0i} = a^2 \, \partial_i B \; , \quad
g_{ij} = a^2 \, \delta_{ij} \; .
\end{equation}
Due to the assumption that the photon field $A_\mu$ has no intrinsic background values and to the fact that it enters the action only quadratically, the effects of the produced fields on the curvature perturbations are formally second-order in the perturbative expansion. To focus on their effects, we therefore treat all the scalar modes as second order and the gauge field as first order, namely,
\begin{equation}
\delta\phi \rightarrow \delta_2 \phi \; , \quad
\delta\chi \rightarrow \delta_2 \chi \; , \quad
\Phi \rightarrow \Phi_2 \; , \quad
B \rightarrow B_2 \; , \quad
A_\mu \rightarrow \delta_1 A_\mu \; ,
\end{equation}
where the subscripts $(1,2)$ are the perturbative orders. Noting this, we suppress them in the following calculations without ambiguity. Expanding the action \eqref{Model Action} up to second order, we understand that the variations with respect to $\Phi$ and $B$ only provide constraint equations. Thus the true equations of motion are those of $\delta\phi$, $\delta\chi$ and $A_\mu$. The one for $A_\mu$ is given in \eqref{EoM of A}, and those for $\delta\phi$ and $\delta\chi$ are found as
\begin{eqnarray}
&& \left( \partial_{\eta}^2 - \nabla^2 \right) Q_i + {\cal M}_{ij}^2 Q_j = {\cal S}_i^{d} + {\cal S}_i^{g} \; ,
\label{eom-dphidchi}
\end{eqnarray} 
where $Q_i = \{ a \, \delta\phi ,\, a \, \delta\chi \}$, ${\eta}$ is the conformal time, and
\begin{eqnarray}
{\cal M}_{ij}^2 &=&  - \frac{a''}{a} \delta_{ij} + a^2 V_{,ij} + \left( 3 - \frac{\phi_k' \phi_k'}{2 M_p^2 {\cal H}^2} \right) \frac{\phi_i' \phi_j'}{M_p^2} + \frac{a^2}{M_p^2 {\cal H}} \left( \phi_i' V_{,j} + \phi_j' V_{,i} \right) \; ,
\nonumber \\
{\cal S}_i^{d} & = & a^3 \, \frac{\partial_\chi I}{I} \, \left( \bm{E}^2 - \bm{B}^2 \right) 
\left(
\begin{array}{c}
0 \\ 1 
\end{array}
\right)_i \; ,
\nonumber\\
{\cal S}_i^{g} & = & - \frac{a^3 \phi_i'}{2 M_{\rm Pl}^2 {\cal H}} \,
\left\{
\frac{\bm{E}^2 + \bm{B}^2}{2}
- \frac{{\cal H}^2}{a^4 \phi_i'{}^2} \, \nabla^{-2} \partial_{\eta} \left[ \frac{a^4 \phi_i'{}^2}{{\cal H}^2} \, \bm{\nabla} \cdot \left( \bm{E} \times \bm{B} \right) \right]
\right\} \; ,
\label{source-app}
\end{eqnarray}
with $\phi_i = \{ \phi_0 , \, \chi_0 \}$, prime denoting derivatives with respect to ${\eta}$, and ${\cal H} \equiv a'/a$.
The first, second and third equations of \eqref{source-app} correspond, respectively, to the mass mixing, the direct coupling between the photon and $\chi$, and the gravitational interaction. Here we have defined 
the electric and magnetic fields as
\begin{equation}
\bm{E} \equiv - \frac{\langle I \rangle}{a^2} \, \partial_{\eta} \bm{A} \; , \quad
\bm{B} \equiv \frac{\langle I \rangle}{a^2} \, \bm{\nabla} \times \bm{A}
\end{equation}
where the bracket denotes background values. The equations of motion, \eqref{eom-dphidchi}, together with \eqref{source-app}, are exact at this order.

Since we are interested in the regime of parameters where the electric contribution is dominant over the magnetic one, we neglect all the terms coming from the latter, except for the one in the term of Poynting vector $\bm{E} \times \bm{B}$. Also in this regime the electric field is always monotonically increasing both during and after inflation until reheating, and thus we solve \eqref{eom-dphidchi} during the period of inflaton oscillation.
To utilize the Green function method, we first find the homogeneous solution of \eqref{eom-dphidchi}. Fourier-transforming $\delta\phi$ and $\delta\chi$ as
\begin{equation}
\delta\phi({\eta}, \bm{x}) = \int \frac{d^3k}{(2\pi)^3} \, {\rm e}^{i \, \bm{k} \cdot \bm{x}} \, \frac{\varphi_{\bm k} ({\eta})}{a({\eta})} \; , \quad
\delta\chi({\eta}, \bm{x}) = \int \frac{d^3k}{(2\pi)^3} \, {\rm e}^{i \, \bm{k} \cdot \bm{x}} \, \frac{{\cal X}_{\bm k} ({\eta})}{a({\eta})} \; ,
\end{equation}
and $A_i$ as in \eqref{introduction of creation/annihilation operator}, the homogeneous equations of \eqref{eom-dphidchi} can be approximated as,%
\footnote{Here we treat the mass mixing terms in \eqref{eom-dphidchi} perturbatively, and thus they do not enter in the free equations of motion.}
\begin{equation}
\varphi_{\bm k}'' + \left( k^2 + a^2 V_{\phi\phi} \right) \varphi_{\bm k } \simeq 0 \; , \quad
{\cal X}_{\bm k}'' + \left( k^2 - \frac{a''}{a} \right) {\cal X}_{\bm k} \simeq 0 \; .
\label{eom-free}
\end{equation}
Since it suffices to consider the period of inflaton oscillation for our current purpose, we solve these equations in this era. 
For later use, we remind readers that for a given equation of state of the universe the scale factor is related to conformal time as $a= \frac{2}{1+3w} \frac{1}{{\eta} H} \propto t^{2/3(1+w)} \propto {\eta}^{2/(1+3w)}$.
Defining the Fourier transform of the source terms in \eqref{source-app} as
$\hat{\cal S}_i^{d/g} \equiv \int d^3x \, {\rm e}^{-i \bm{k} \cdot \bm{x}} \, {\cal S}_i^{d/g}$,
they can be approximated as,
\begin{eqnarray}
\hat{\cal S}_i^d \left( {\eta} , \bm k \right) &\simeq&
2 \, a^3 \, \frac{\partial_\chi I}{I} \,
\delta\hat{\rho}_E \,
\left(
\begin{array}{c}
0 \\ 1
\end{array}
\right)_i \; ,
\nonumber\\
\hat{\cal S}_i^g \left( {\eta} , \bm k \right) & \simeq &
-\frac{a^3 \phi_i'}{2 M_{\rm Pl}^2 {\cal H}} \,
\left[
 \delta\hat{\rho}_E
- \frac{i \, k_j}{k^2} \, \frac{{\cal H}^2}{a^4 \phi_i'{}^2} \, \partial_{\eta} \left( \frac{a^4 \phi_i'{}^2}{{\cal H}^2} \, \delta\hat{P}_j \right)
\right] \; ,
\label{source-FT}
\end{eqnarray}
where the electric energy density and the Poynting vector are
\begin{eqnarray}
\delta \hat{\rho}_E \left( {\eta} , \bm{k} \right) &\equiv& \frac{1}{2} \int \frac{d^3p}{(2\pi)^3} \, \hat{E}_i \left( {\eta} , \bm p \right) \, \hat{E}_i \left( {\eta} , \bm{k} - \bm{p} \right) \; ,
\nonumber\\
\delta\hat{P}_i \left( {\eta} , \bm{k} \right) &\equiv&
\int \frac{d^3p}{(2\pi)^3} \, \epsilon_{ijk} \, \hat{E}_j \left( {\eta} , \bm p \right) \, \hat{B}_k \left( {\eta} , \bm{k} - \bm{p} \right) \; ,
\end{eqnarray}
with $\hat{E}_i$ and $\hat{B}_i$ being the Fourier-transformed $\bm{E}$ and $\bm{B}$ fields, respectively. Note that although the term of the Poynting vector in the source $\hat{\cal S}_i^g$ appears divergent in the limit $k \rightarrow 0$, physical spectra do not suffer IR divergence, which we shall show in Subsection \ref{subapp:total}. 

To compute the curvature perturbation resulting from the produced photon field, we have the relation
\begin{equation}
\zeta = - \frac{H}{\dot\rho} \, \delta\rho \; ,
\label{rel-zetarho}
\end{equation}
in the spatially flat gauge. The density perturbation $\delta\rho$ consists of three contributions, from the bare electric energy density $\delta\rho_E$ (with negligible magnetic component), and the density perturbations of spectator and inflaton fields, $\delta\rho_\chi$ and $\delta\rho_\phi$, respectively, sourced by the electromagnetic fields. Their formal expressions are written as, up to the relevant orders,
\begin{equation}
\delta\rho_E = \frac{\bm{E}^2}{2} \; , \quad
\delta\rho_\chi = \dot\chi_0 \, \delta\dot\chi + U_{,\chi} \, \delta\chi \; , \quad
\delta\rho_\phi = \dot\phi_0 \, \delta\dot\phi + V_{,\phi} \, \delta\phi \; .
\label{density-pert-all}
\end{equation}
In the following subsections, we compute the contributions from the photon field to $\delta\chi$ and $\delta\phi$ separately, and then the total one to the curvature perturbation.
For concrete calculation, we hereafter set $w=0$ during the phase of inflaton oscillation, that is the inflaton oscillates around its potential $V(\phi) = m_\phi^2 \phi^2 / 2$.

\subsection{Contributions to spectator field perturbation}

The equation of motion for the spectator field perturbation during inflaton oscillation can be obtained from the homogeneous part \eqref{eom-free} together with the source from the electromagnetic field, \eqref{source-FT}, and reads
\begin{equation}
{\cal X}_{\bm k}'' + \left[ k^2 - \frac{2}{{\eta}^2}\right] {\cal X}_{\bm k} \simeq 
-2 \, n \, \frac{a^3 H}{\dot\chi_0} \, \delta\hat{\rho}_E
\; ,
\label{eom-X-source}
\end{equation}
where we have set $w=0$ and used the relation $\partial_\chi I / I = \dot I / (\dot\chi_0 I) = - n H / \dot\chi_0$.
Here we neglect the Planck suppressed operators, as they make only sub-leading contributions.
Note that when the mass of the spectator field is light, i.e. $U_{,\chi\chi} \ll H^2$, the background $\chi_0$ follows the approximate time dependence $\frac{9}{2} H \dot\chi_0 \simeq - U_{,\chi}$ in the matter-dominated universe. We assume this ``slow roll'' of $\chi_0$, and under this assumption, the time evolution of $\dot\chi_0$ is approximately $\dot\chi_0 \propto H^{-1}$.

The Green function associated with ${\cal X}$ can be found by equating the homogeneous part of \eqref{eom-X-source} to $\delta \left( {\eta} - {\eta}' \right)$, giving
\begin{equation}
G_k^\chi \left( {\eta} , {\eta}' \right) = \Theta\left( {\eta} - {\eta}' \right) \, \frac{\pi}{2} \sqrt{{\eta} {\eta}'} \left[
Y_{3/2} \left( \vert k {\eta}\vert \right) \, J_{3/2} \left( \vert k {\eta}'\vert \right) - J_{3/2} \left( \vert k {\eta}\vert \right) \, Y_{3/2} \left( \vert k {\eta}'\vert \right)
\right] \; ,
\end{equation}
where $\Theta$ is the Heaviside step function.
The part of the solution due to the electromagnetic source can then be solved as
\begin{equation}
{\cal X}_{\bm k}^{\em} \left( {\eta} \right) = -2n  \int_{-\infty}^\infty d{\eta}' \, G_k^\chi \left( {\eta} , {\eta}' \right) \, \left[ \frac{a^3H}{\dot\chi_0} \right]_{{\eta}'} \, \delta\hat\rho_E \left( {\eta}' , \bm{k} \right) \; ,
\label{X-sol-formal}
\end{equation}
where superscript $\em$ denotes a quantity sourced by the electromagnetic field, and subscript ${\eta}'$ indicates a quantity evaluated at time $\eta'$.
To evaluate the time integral, we are only interested in super-horizon modes during the period of inflaton oscillation, as the modes inside the horizon damp away quickly and leave negligible contributions. In this limit, we have
\begin{eqnarray}
G_k^\chi \left( {\eta} , {\eta}' \right) & \simeq & \frac{\Theta\left( {\eta} - {\eta}' \right)}{3}
\sqrt{ {\eta} {\eta}'}
\left[ \left( \frac{{\eta}}{{\eta}'} \right)^{3/2} - \left( \frac{{\eta}'}{{\eta}} \right)^{3/2} \right] \; ,
\end{eqnarray}
$a^3H / \dot\chi_0 = {\rm const.}$, and $\delta\hat\rho_E \propto a^{2n-4} \propto
{\eta}^{4n-8}$. 
Then \eqref{X-sol-formal} is evaluated to be
\begin{equation}
{\cal X}_{\bm k}^{\em} \left( {\eta} \right) \simeq
a({\eta}) \, \frac{-2n}{\left( n-2 \right) \left( 4n-5 \right)} \, \frac{\delta\hat\rho_E ( {\eta} , \bm k )}{\dot\chi_0 H} \; .
\label{X-source-sol}
\end{equation}
Notice that $\dot\chi_0 H$ is constant, and therefore the time dependence of the physical mode follows that of the electric energy density, i.e. ${\cal X}^{\em} / a \propto \delta\hat\rho_E$.

Using \eqref{density-pert-all} and recalling the relation $U_{,\chi} \simeq - 9 H \dot\chi_0/2$ (see below \eqref{eom-X-source}), we obtain, for the sourced $\delta\rho_\chi$,
\begin{equation}
\delta\hat\rho_\chi^{\em} \simeq \frac{4n-17}{2} \, \dot\chi_0 H \delta\hat\chi^s \simeq
\frac{-n \left( 4n-17 \right)}{\left( n-2 \right) \left( 4n - 5 \right)} \, \delta\hat\rho_E \; ,
\label{drho-chi}
\end{equation}
where hat denotes operators in the Fourier space, and superscript $s$ the sourced mode.

\subsection{Contributions to inflaton perturbation}

We now turn to the contributions to inflaton perturbations from the produced electromagnetic fields. In this computation, we switch to using physical time $t$ instead of conformal one ${\eta}$. As it becomes clear below, we need to take into account the contributions from sub-leading orders in ${\cal O}(H/m_\phi)$. For this reason, we include up to the first order in $H/m_\phi$ and find
\begin{eqnarray}
\phi_0 &\cong& \phi_e \left( \frac{a_e}{a} \right)^{3/2} \cos\theta(t) \; , \nonumber\\
\dot\phi_0 &\cong& - m_\phi \, \phi_e \left( \frac{a_e}{a} \right)^{3/2} \left[ \sin\theta(t) + \frac{3H_{\rm inf}}{2m_\phi} \left( \frac{a_e}{a} \right)^{3/2} \cos\theta(t) \right] \; ,\nonumber\\
H &\cong& H_{\rm inf} \left( \frac{a_e}{a} \right)^{3/2} \left[ 1+ \frac{3H_{\rm inf}}{4m_\phi} \left( \frac{a_e}{a} \right)^{3/2} \sin 2\theta(t) \right] \; ,
\end{eqnarray}
where $\theta(t) = m_\phi (t-t_e)$, $\phi_e = \sqrt{6} \, M_{\rm Pl} H_{\rm inf} / m_\phi$, and subscripts $e$ and ${\rm inf}$ denote values at the end of and during inflation, respectively. Note that $a/a_0 = (3H_{\rm inf} \, t/2)^{2/3}$ up to this order.

The source term for $\delta\phi$ consists of two contributions, one from the gravitational interaction with the gauge field, $\hat S_\phi^g$, and the other from the mass mixing with $\delta\chi$, ${\cal M}_{\phi\chi}$. Writing up to the terms one-order suppressed by $H/m_\phi$, we find%
\footnote{One can show that the next-to-leading order contribution to ${\cal X}_{\bm k}^s$ is $\mathcal{O}(H^3/m_\phi^3)$ and thus negligible.}
\begin{eqnarray}
\hat S_\phi^g &\cong& \sqrt{\frac{3}{2}} \, \frac{a^3}{M_{\rm Pl}}
\bigg\{
- 2 i m_\phi \frac{a k_i}{k^2} \, \delta\hat{P}_i \, \cos\theta
+ \delta\hat\rho_E \, \sin\theta
\nonumber\\ && \qquad
- i H_{\rm inf} \left( \frac{a_e}{a} \right)^{3/2} \frac{a k_i}{k^2} \, \delta\hat{P}_i \left[ \left( 2n - \frac{7}{4} \right) \sin\theta - \frac{9}{4} \sin3\theta \right]
\bigg\} \; ,
\nonumber\\
{\cal M}_{\phi\chi}^2 &\cong& \sqrt{6} \, a^2 \frac{\dot\chi_0}{M_{\rm Pl}} \, m_\phi \left[ 
\cos\theta + \frac{9H_{\rm inf}}{8m_\phi} \left( \frac{a_e}{a} \right)^{3/2} \left( 3 \sin\theta - \sin 3\theta \right) 
\right] \; .
\label{phi-sources}
\end{eqnarray}
The first terms in both expressions appear dominant; however, we will see that they contribute to the curvature perturbation only in the same order as the other terms (and that is why we need to include the apparently sub-leading terms). In the contribution from $\delta\chi$, we concentrate on the effect from the gauge field, which is given by \eqref{X-source-sol}. Therefore we can express the total source for the inflation perturbation from the gauge field as
\begin{equation}
\hat{S}_\phi \equiv \hat{S}_\phi^g - {\cal M}_{\phi\chi}^2 {\cal X}_{\bm k}^{\em} \; ,
\label{phi-source-tot}
\end{equation}
where $\hat{S}_\phi^g$ and ${\cal M}_{\phi\chi}^2$ are given in \eqref{phi-sources}, and ${\cal X}_{\bm k}^s$ in \eqref{X-source-sol}.

The equation of motion for the inflaton perturbation during inflaton oscillation can be obtained from the homogeneous part in \eqref{eom-free} together with the source \eqref{phi-source-tot}. In physical time, we have%
\footnote{
We neglect the next-to-leading order in ${\cal M}_{\phi\phi}$, coming from the $\phi_0' V_{,\phi}$ term, which might in principle contribute to the homogeneous solution for $\varphi_{\bm k}$ and therefore to its Green function. This would modify our result {\it at most} by an ${\cal O}(1)$ numerical factor, and thus our main message would not be altered.
}
\begin{equation}
\left( \partial_t^2 + \frac{k^2}{a^2} + m_\phi^2 \right) \left( a^{1/2} \varphi_{\bm k} \right) \simeq a^{-3/2} \hat{S}_\phi \; .
\end{equation}
This equation can be formally solved for the sourced part of the solution as
\begin{eqnarray}
a^{1/2} \varphi_{\bm k}^{\em} (t) &=&  \int dt' \, G_k^\phi \left( t, t' \right) \, a^{-3/2} (t') \, \hat{S}_\phi \left( t' , \bm{k} \right) \; ,
\nonumber\\
\partial_t \left[ a^{1/2} \varphi_{\bm k}^{\em} (t) \right] &=&  \int dt' \, \partial_t G_k^\phi \left( t, t' \right) \, a^{-3/2} (t') \, \hat{S}_\phi \left( t' , \bm{k} \right) \;
\label{phi-formal}
\end{eqnarray}
Focusing on the super-horizon modes, i.e., $k/a \ll m_\phi$, the Green function $G_k^\phi(t,t')$ is found as
\begin{equation}
G_k^\phi(t,t') = \Theta(t-t') \, \frac{\sin\left[ m_\phi(t-t') \right]}{m_\phi} \; .
\end{equation}
Evaluating the time integrals in \eqref{phi-formal} for the period of inflaton oscillation, during which the electromagnetic field evolves as $\delta\hat\rho_E \propto a^{2n-4}$ and $\delta\hat P_i \propto a^{2n - 7/2}$, we find
\begin{eqnarray}
a^{1/2} \varphi_{\bm k}^{\em} &\simeq&
A_1 \left( \frac{t}{t_e} \right)^{\frac{4n+1}{3}} \left( \frac{m_\phi}{H_{\rm inf}} \sin\theta + \frac{4n+1}{4} \, \frac{t_e}{t} \cos\theta \right)
- A_2 \left( \frac{t}{t_e} \right)^{\frac{4n-2}{3}} \cos\theta
\; ,
\nonumber\\
\frac{\partial_t \left( a^{1/2} \varphi_{\bm k}^{\em} \right)}{m_\phi} &\simeq&
A_1 \left( \frac{t}{t_e} \right)^{\frac{4n+1}{3}} \left( \frac{m_\phi}{H_{\rm inf}} \cos\theta + \frac{4n+1}{4} \, \frac{t_e}{t} \sin\theta \right)
+ A_2 \left( \frac{t}{t_e} \right)^{\frac{4n-2}{3}} \sin\theta \; ,
\nonumber\\
\label{phi-dphi}
\end{eqnarray}
where
\begin{eqnarray}
A_1 &\equiv& \frac{\sqrt{6} \, a_e^{3/2}}{M_{\rm Pl} m_\phi H_{\rm inf} (4n+1)} \left[
\frac{2n}{(n-2)(4n-5)} \, \delta\hat\rho_E(t_e) - i \, \frac{a_e H_{\rm inf} k_i}{k^2} \, \delta\hat P_i (t_e)
\right] \; ,
\nonumber\\
A_2 &\equiv& \frac{\sqrt{6} \, a_e^{3/2}}{M_{\rm Pl} m_\phi H_{\rm inf} (4n-2)} \left[
\frac{8n^2+n+20}{4(n-2)(4n-5)} \, \delta\hat\rho_E (t_e) - i \left( n - \frac{7}{8} \right) \frac{a_e H_{\rm inf} k_i}{k^2} \, \delta\hat P_i (t_e)
\right] \; .
\nonumber\\
\end{eqnarray}
Using the definition of $\delta\rho_\phi$ in \eqref{density-pert-all}, one can easily see that the dominant terms of the equations in \eqref{phi-dphi} cancel out. This is because the would-be leading (dangerous) contributions in $\delta\phi$ interfere destructively with the oscillating background $\phi_0$. We thus obtain the time-averaged perturbation of inflaton energy density sourced by the EM field,
\begin{equation}
\overline{\delta\hat\rho_\phi^{\em}} \simeq 
\frac{3}{2n-1} \left[-
\frac{8n^2+n+20}{4(n-2)(4n-5)} \, \delta\hat\rho_E (t) + i \left( n - \frac{7}{8} \right) \frac{a H k_i}{k^2} \, \delta\hat P_i (t)
\right] \; ,
\label{drho-phi}
\end{equation}
up to the actual leading order, where bar denotes the time average.

\subsection{Total energy density perturbation}
\label{subapp:total}

The total energy density perturbation is a simple summation of density perturbations of electric, spectator and inflaton, i.e. $\delta\rho_{\rm tot} = \delta\rho_E + \delta\rho_\chi + \delta\rho_\phi$. Using \eqref{drho-chi} and \eqref{drho-phi}, we obtain the total perturbation originated from the gauge field,
\begin{equation}
\delta\hat\rho_{\rm tot}^{\em} \simeq
\frac{8n^2+61n-100}{4(n-2)(2n-1)(4n-5)} \, \delta\hat\rho_E
+ i \, \frac{3(8n - 7)}{8(2n-1)} \, \frac{a H k_i}{k^2} \, \delta\hat P_i \; ,
\label{drhotot}
\end{equation}
where hat denotes an operator in the Fourier space.
Before proceeding to compute correlation functions, let us comment on convergence of the $\delta\hat P_i$ term. At the operator level, it is straightforward to see
\begin{eqnarray}
i \, \frac{k_i}{k^2} \, \delta\hat P_i (\bm k)
&=&
\frac{I^2}{a^4} \int \frac{d^3p}{(2\pi)^3} \bigg\{
\frac{p}{2k} \, \hat{\bm k} \cdot \hat{\bm p} \left[ \hat A_i'(\bm k - \bm p) \, \hat A_i(\bm p) - \hat A_i'(\bm p) \, \hat A_i(\bm k - \bm p) \right]
\nonumber\\ && \qquad\qquad\qquad\quad
- \left( \hat k_i  \hat k_j - \frac{\delta_{ij}}{2} \right) \, \hat A_i' (\bm p) \, \hat A_j (\bm k - \bm p)
\bigg\} \; ,
\label{dPi-operator}
\end{eqnarray}
where we have sent the integration variable $\bm p \rightarrow \bm k - \bm p$. A glance at the first term in the curly parentheses of \eqref{dPi-operator} seems to hint that this quantity diverges in the IR limit, $k \rightarrow 0$. We now show that this is not the case. By decomposing $\hat A_i (t, \bm p) = \epsilon_i^\lambda \left( \hat {\bm p} \right) A_\lambda (t, \bm p)$, 
we see, in the limit $k \rightarrow 0$,
\begin{equation}
\hat A_i'(\bm k - \bm p) \, \hat A_i(\bm p) - \hat A_i'(\bm p) \, \hat A_i(\bm k - \bm p) \rightarrow
A_{\lambda}'(- \bm p) \, A_\lambda(\bm p) - A_\lambda'(\bm p) \, A_{\lambda}(- \bm p) \; .
\end{equation}
From this, one can easily show that as long as the mode function in $A_\lambda$ is real up to a constant phase (which is generally true once modes become classical), the right-hand side vanishes. Therefore, as long as we consider semi-classical (statistical) quantities, we can quite generally conclude that $k_i \delta\hat P_i / k^2$ stays finite in the limit $k \rightarrow 0$.

The two-point correlation function of $\delta\hat\rho_{\rm tot}^\em$ consists of the following three contributions:
\begin{eqnarray}
&& \left\langle \delta\hat\rho_E \left( \bm k \right) \delta\hat\rho_E \left( \bm k' \right) \right\rangle = \frac{\delta^{(3)} \left( \bm{k} + \bm{k}' \right)}{2} \int d^3p
\left[ 1 + \left( \hat{\bm p} \cdot \widehat{\bm k - \bm p} \right)^2 \right]
\big\vert E(p) \big\vert^2  \, \big\vert E( \vert \bm k - \bm p \vert ) \big\vert^2 \; ,
\nonumber\\
&& \left\langle i \frac{k_i}{k^2} \delta\hat P_i \left( \bm k \right) \, i \frac{k_j'}{k'{}^2} \delta\hat P_j \left( \bm k' \right) \right\rangle = \frac{\delta^{(3)} ( \bm k + \bm k')}{k^2} \int d^3p
\nonumber\\ && \qquad\qquad
\times \Big\{
\left[ \left( \hat{\bm k} \cdot \hat{\bm p} \right)^2 + \left( \hat{\bm k} \cdot \widehat{\bm{k} - \bm{p}} \right)^2 \right] \big\vert E (p) \big\vert^2 \, \big\vert B ( \vert \bm k - \bm p \vert ) \big\vert^2
\nonumber\\ && \qquad\qquad\quad
+ 2 \left( \hat{\bm k} \cdot \hat{\bm p} \right) \left( \hat{\bm k} \cdot \widehat{ \bm k - \bm p} \right) \, E(p) \, B^*(p) \, B(\vert \bm k - \bm p \vert) \, E^*( \vert \bm k - \bm p \vert )
\Big\} \; ,
\nonumber\\
&& \left\langle i \frac{k_i}{k^2} \delta\hat P_i \left( \bm k \right) \, \delta\hat\rho_E \left( \bm k' \right) \right\rangle = - \frac{\delta^{(3)} \left( \bm{k} + \bm{k}' \right)}{k} \int d^3p
\nonumber\\ && \qquad\qquad
\times \left[ \hat{\bm k} \cdot \hat{\bm p} + \left( \hat{\bm k} \cdot \widehat{ \bm k - \bm p} \right) \left( \hat{\bm p} \cdot \widehat{\bm k - \bm p} \right) \right]
B (p) \, E^*(p) \, \big\vert E (\vert \bm k - \bm p \vert) \big\vert^2 \; ,
\label{twopoints-EM}
\end{eqnarray}
where
\begin{equation}
E(p) \equiv - \frac{I}{a^2} {\cal A}'(p) \; , \quad 
B(p) \equiv \frac{I}{a^2} \, p \, {\cal A}(p) \; , \quad
{\cal A}(p) \equiv {\cal A}_+(p) = {\cal A}_-(p) \; .
\end{equation}
We are interested in the scales relevant to CMB observations, i.e. the external momentum $k \sim k_{\rm CMB}$, while the phase space momentum $\vec p$ is peaked around $p \sim a_i H_{\rm inf} \gg k_{\rm CMB}$, for our current interest in magnetogenesis with coherent length $L \lesssim 1 \; {\rm Mpc}$. In this limit, the first quantity in \eqref{twopoints-EM} simplifies to
\begin{equation}
\left\langle \delta\hat\rho_E \left( \bm k \right) \delta\hat\rho_E \left( \bm k' \right) \right\rangle \simeq \delta^{(3)} \left( \bm{k} + \bm{k}' \right) \int d^3p \,
\big\vert E(p) \big\vert^4 \; .
\label{drhoEdrhoE-1}
\end{equation}
The expressions for $E(p)$ and $B(p)$ outside horizon during the period of inflaton oscillation can be approximated as
\begin{equation}
E\left( {\eta} , p \right) \simeq
C_2 \!\left( \frac{p}{k_i} \right) \frac{2^{n + \frac{1}{2}} \, \Gamma \left( n+\frac{1}{2} \right)}{\pi \, p^{3/2}}
\left( \frac{a_e H_{\rm inf}}{p} \right)^n 
\left( \frac{a_e H_{\rm inf}}{aH} \right)^{2n}
\left( \frac{p}{a} \right)^2
\, , \;
B\left( {\eta} , p \right) \simeq
- \frac{p{\eta}}{4n+1} \,
E \left( {\eta} , p \right) \; ,
\label{EB-approx}
\end{equation}
where subscripts $e$ and ${\rm inf}$ denote values at the end of and during inflation, respectively, and we recall the definition of $C_2$ in eq.~\eqref{fullC2}
\begin{equation}
C_2 (x) = \frac{\pi}{2 \sqrt{2}} \sqrt{x} \left[ J_{n + \frac{1}{2}} \left( x \right) + i \, J_{n - \frac{1}{2}} \left( x \right) \right] \; .
\end{equation}
Plugging \eqref{EB-approx} into \eqref{drhoEdrhoE-1}, we find
\begin{equation}
\left\langle \delta\hat\rho_E \left( \bm k \right) \delta\hat\rho_E \left( \bm k' \right) \right\rangle \big\vert_{t=t_r} \simeq \delta^{(3)} \left( \bm{k} + \bm{k}' \right) 
\frac{2^{4(n+1)} \, \Gamma^4\left( n + \frac{1}{2} \right)}{\pi^3 \, a_r^8}
\left( a_i H_{\rm inf} \right)^5
\left( \frac{a_e}{a_i} \right)^{4n}
\left( \frac{a_e H_{\rm inf}}{a_r H_r} \right)^{8n}
G_n^{(1)} \; ,
\label{drhoEdrhoE}
\end{equation}
where the quantity is evaluated at the time of reheating, denoted by subscripts $r$, and
\begin{equation}
G_n^{(1)} \equiv \int_0^\infty dz \,
\vert C_2 \left( z \right) \vert^4 \, z^{-4n+4} \; ,
\label{def-Gn}
\end{equation}
The lower and upper bounds of the integral in \eqref{def-Gn} should in principle be numbers of ${\cal O} \left( \frac{a_r H_r}{a_i H_{\rm inf}} \right)$ and ${\cal O} \left( \frac{k}{a_i H_{\rm inf}} \right)$, respectively. However, the integrand is peaked around $z \sim {\cal O}(1)$, which is well within the domain of integration, and sending the lower and upper bounds respectively to $0$ and $\infty$ is a good approximation. We compute the integral \eqref{def-Gn} numerically, and the result can be fitted as, within the domain $2<n<10$,
\begin{equation}
G_n^{(1)} \simeq \exp\left( 5.27 - 2.34 n - 0.821 n^2 + 0.0240 n^3 \right) \; ,
\end{equation}
with an error of $\lesssim 1 \, \%$.

For the second and third quantities in \eqref{twopoints-EM}, the would-be leading-order terms vanish in the expansion for small $k/p$. The next would-be leading order also vanishes, since they are proportional to odd orders in $\hat k \cdot \hat p$ and thus give zero after the angular integration. This explicitly demonstrates that the $\delta\hat P_i$ terms do not suffer IR divergence, as discussed at the beginning of this subsection.
Then the actual leading-order contributions are
\begin{eqnarray}
\left\langle i \frac{k_i}{k^2} \delta\hat P_i \left( \bm k \right) \, i \frac{k_j'}{k'{}^2} \delta\hat P_j \left( \bm k' \right) \right\rangle &\simeq&
\delta^{(3)} ( \bm k + \bm k') \,
\frac{2^{4n+2} \pi \, \Gamma^4\left( n + \frac{1}{2} \right)}{15 \left( 4n+1 \right)^2 a^8} \, 
\frac{\left( a_i H_{\rm inf} \right)^5}{\left( a H \right)^2}
\left( \frac{a_e}{a_i} \right)^{4n}
\left( \frac{a_e H_{\rm inf}}{aH} \right)^{8n}
G_n^{(2)} \; ,
\nonumber \\
\left\langle i \frac{k_i}{k^2} \delta\hat P_i \left( \bm k \right) \, \delta\hat\rho_E \left( \bm k' \right) \right\rangle &\simeq&
\delta^{(3)} ( \bm k + \bm k') \,
\frac{2^{4n+1} \pi \, \Gamma^4\left( n + \frac{1}{2} \right)}{3 \left( 4n+1 \right) a^8} \,
\frac{\left( a_i H_{\rm inf} \right)^{5}}{a H}
\left( \frac{a_e}{a_i} \right)^{4n}
\left( \frac{a_e H_{\rm inf}}{aH} \right)^{8n} 
G_n^{(3)} \; ,
\nonumber\\
\label{PP-Prho}
\end{eqnarray}
where
\begin{eqnarray}
G_n^{(2)} & \equiv &  \int_0^\infty dz \, z^{6-4n} 
\left[ J_{n+1/2}^2 (z) + J_{n-1/2}^2 (z) \right]
\bigg[ \left( 6 n - 1 \right) J_{n+1/2}^2 (z) - J_{n-1/2}^2 (z) \bigg]  \; ,
\nonumber \\
G_n^{(3)} & \equiv & \int_0^\infty dz \,
z^{6-4n}
\left[ J_{n+1/2}^2 ( z ) + J_{n-1/2}^2 (z) \right]
\left[ \left( 2 n -1 \right) J_{n+1/2}^2 (z) - J_{n-1/2}^2 (z) \right] \; .
\nonumber\\
\end{eqnarray}
These functions can be fitted as, in the regime $2<n<10$,
\begin{eqnarray}
G_n^{(2)} &\simeq& \exp\left( 5.86 - 2.34 n - 0.820 n^2 + 0.0240 n^3 \right) \; ,
\nonumber\\
G_n^{(3)} &\simeq& \exp\left( 3.46 - 2.33 n - 0.821 n^2 + 0.0241 n^3 \right) \; ,
\end{eqnarray}
and we hence see $G_n^{(3)} \simeq G_n^{(1)} / 6 \simeq G_n^{(2)} / 11$ to a good agreement.

We are now ready to collect the power spectrum of curvature perturbation $\zeta$, defined as
\begin{equation}
{\cal P}_\zeta \left( k \right)  \, \left( 2 \pi \right)^3 \delta^{(3)} \left( \bm k + \bm k' \right) \equiv \frac{k^3}{2\pi^2} 
\left\langle \hat\zeta \left( \bm k \right) \hat\zeta \left( \bm k' \right) \right\rangle \; ,
\end{equation}
where hat denotes an operator in the Fourier space. Using the relation \eqref{rel-zetarho}, together with the expression \eqref{drhotot} for the total density perturbation, and recalling the background equation $\dot\rho = -3H \rho = - 9 M_{\rm Pl}^2 H^3$, we obtain the sourced power spectrum evaluated at the time of reheating, $t=t_r$,
\begin{eqnarray}
{\cal P}_\zeta^{\em} \big\vert_{t=t_r} &\simeq& 
\frac{2^{4n} \, \Gamma^4\left( n + \frac{1}{2} \right)}{81 \pi^8}
\left[
\gamma_n^2 G_n^{(1)}
+ \frac{\pi^4 \lambda_n^2 G_n^{(2)}}{60 \left( 4n+1 \right)^2}
+ \frac{\pi^4 \gamma_n \lambda_n G_n^{(3)}}{12 \left( 4n+1 \right)}
\right]
\nonumber\\ && \times
\left( \frac{H_r}{M_{\rm Pl}} \right)^4
\left( \frac{a_i H_{\rm inf}}{a_r H_r} \right)^{5}
\left( \frac{a_e}{a_i} \right)^{4n}
\left( \frac{a_e H_{\rm inf}}{a_r H_r} \right)^{8n} 
\left( \frac{k}{a_r H_r} \right)^3 \; ,
\label{powerzeta-result}
\end{eqnarray}
where
\begin{equation}
\gamma_n \equiv \frac{8n^2+61n-100}{4(n-2)(2n-1)(4n-5)} \; , \quad
\lambda_n \equiv \frac{3(8n - 7)}{8(2n-1)} \; .
\label{def-gamlam}
\end{equation}
Eq.~\eqref{powerzeta-result} is the main result of this appendix. The total power spectrum is the simple sum of the vacuum and sourced modes, denoted respectively by superscripts ${\rm vac}$ and ${\rm em}$, i.e.
\begin{equation}
{\cal P}_\zeta^{\rm tot} = {\cal P}_\zeta^{\rm vac} + {\cal P}_\zeta^{\em} \; .
\end{equation}
Since the sourced spectrum \eqref{powerzeta-result} is strongly scale-dependent, we, at the very least, enforce ${\cal P}_\zeta^{\em} < {\cal P}_\zeta^{\rm vac} \simeq {\cal P}_\zeta^{\rm tot}$. This puts a constraint on the production of magnetic fields in the model of our current consideration. In the main text, we discuss in detail the final magnetic-field strength with this bound on curvature perturbation imposed.


\end{document}